\documentclass[reprint,,twocolumn,color,superscriptaddress,psfig,showpacs,amsmath,amssymb,floatfix,fleqn]{revtex4-1}

\usepackage{epsf}

\usepackage{amssymb}
\usepackage{amsmath}
\usepackage{natbib}
\setcitestyle{square,numbers}
\usepackage{amsthm}
\usepackage{graphicx}% Include figure files
\usepackage{dcolumn}% Align table columns on decimal point
\usepackage{bm}% bold math
\usepackage{color}
\usepackage{subfigure}
\usepackage[subnum]{cases}
\usepackage{float}
\usepackage{epstopdf}

\begin{document}

%\preprint{APS/123-QED}

\title{Design and Performance Analysis of a Chaotic Pseudo Orthogonal Carriers Multi-Access Communication System}

\author{Chao Bai}
\affiliation{Xi'an Key Laboratory of Intelligent Equipment, Xi'an Technological University, Xi'an, 710021, China }
\author{Jun-Liang~Yao}
 \affiliation{Shaanxi Key Laboratory of Complex System Control and Intelligent Information Processing, Xi'an University of Technology, Xi'an, 710048, China}
\author{Yu-Zhe~Sun}
 \affiliation{Shaanxi Key Laboratory of Complex System Control and Intelligent Information Processing, Xi'an University of Technology, Xi'an, 710048, China}
\author{Hai-Peng~Ren}\thanks{Corresponding author:Hai-Peng Ren,renhaipeng@xaut.edu.cn}
  \affiliation{Xi'an Key Laboratory of Intelligent Equipment, Xi'an Technological University, Xi'an, 710021, China }
 \affiliation{Shaanxi Key Laboratory of Complex System Control and Intelligent Information Processing, Xi'an University of Technology, Xi'an, 710048, China}

\date{\today}

\begin{abstract}
A Chaotic Pseudo Orthogonal Carriers Multi-Access (CPOCMA) communication based on Chaotic Pseudo Orthogonal Shape-forming Filter (CPOSF) bank, Chaotic Pseudo Orthogonal Matched Filter (CPOMF) bank and Chaotic Pseudo Orthogonal Correlation Filter (CPOCF) bank is proposed in this work. At the transmitter, the multiple CPOSFs are used to generate pseudo orthogonal signals. It provides a good trade-off between spectrum efficiency and high bit transmission rate. At the receiver, the CPOMF bank and CPOCF bank are used to maximize the Signal-to-Noise Ratio (SNR) and extract the received information from each sub-channel, respectively. The received signal is demodulated by averaging the sampled sequence from the matched filter bank output and sorting the sampling sequence from the CPOCF bank output to recover the transmitted information bits. The proposed CPOCMA communication system not only offers multiuser access with high reliability and high data transmission rate, but also achieves higher spectrum efficiency. Analytical Bit Error Rate (BER) expression is derived. The proposed communication system performance has been evaluated in Additive White Gaussian Noise (AWGN) channel and wireless channel by both numerical simulations and experiments based on a Wireless open-Access Research Platform (WARP), the results show the effectiveness and the superiority of the proposed method.
\end{abstract}

\maketitle

\section{Introduction}
It is predicted that, by 2025, more than 50 billion Internet of Things (IoT) devices will be connected to Internet-enabled systems \cite{Ansere2020_Optimal}. The spectral and power efficiency, low complexity and interference resistance are top requirements for wireless communication in IoT applications \cite{Palattella2016_Internet}, which bring great challenges for traditional communication systems. The chaos-based communications with low cost and low power consumption have been considered to accommodate the demand of the IoT devices in the past decade \cite{Kolumban2019_New}, which has been successfully applied in commercial fiber optic channels \cite{Apostolos2005_Chaos} and in IEEE local network communication standard \cite{IEEEstd2012_Wireless}.

Chaos-based digital communications systems with high Bit Transmission Rate (BTR) are one of the most focusing research fields \cite{Kaddoum2016_Wireless,Cai2016_A,Cai2020_Multicarrier,Bai2020_Double}. Among the proposed communications schemes, Frequency-Division Multiplexing (FDM) technique is a hot research topic in wireless communications systems because the advantages of frequency selective channel resistance and the jamming mitigation. A typical scheme Multi-Carrier Differential Chaos Shift Keying (MC-DCSK) was proposed in \cite{Kaddoum2013_Design} which transmitted chaotic reference signal over a pre-defined subcarrier frequency, at the same time, multiple modulated data streams were transmitted over the other subcarriers. The MC-DCSK improves energy efficiency and increases BTR, but it demands large bandwidth. The subcarrier power optimization of MC-DCSK was performed to improve the Bit Error Rate (BER) performance in \cite{Dai2015_Optimal}, IFFT/FFT operations was used in \cite{Kaddoum2016_Design} instead of parallel matched filters to reduce the implementation complexity and computation cost of MC-DCSK. The shared frequencies were used to transmit the information bits in \cite{Xu2010_A} to save bandwidth and remove the Radio Frequency (RF) delay line required by the traditional DCSKs. The Repeated Spreading Sequence MC-DCSK (RSS-MC-DCSK) in \cite{Nguyen2018_Multi} used the repeated symbol sequences to increase the multipath propagation resistance and to decrease the BER in complex channels. The Multi-Carrier Chaos Shift Keying (MC-CSK) in \cite{Yang2017_Multi} and Sub-carrier Allocated MC-DCSK (SA-MC-DCSK) in \cite{Yang2018_Multi} effectively decreased the BER at the cost of high implementation complexity and low spectrum efficiency, respectively. However, most existing chaos-based communication systems generally employ the chaotic signals generated by discrete Logistic
\cite{Kaddoum2016_Wireless,Cai2016_A,Cai2020_Multicarrier,Bai2020_Double,Kaddoum2013_Design,Dai2015_Optimal,Kaddoum2016_Design,Xu2010_A,Nguyen2018_Multi,Yang2017_Multi,Yang2018_Multi,Zhang2020_Efficient,Tarable2020_Chaos}
or Tent map \cite{Rosa1997_Noise,Li2017_An}, which are used to replace the pseudo-random sequence in most cases. These chaotic signals generated either by discrete map or continuous paradigm systems like Lorenz system \cite{Pappu2020_Simultaneous} are not compatible with the matched filter in the conventional communication systems, moreover, the corresponding modulation schemes usually demonstrate poor spectral efficiency. In response to the growing demand of access terminals, the research demand of the communication systems with higher BTR and higher spectral efficiency are rapid increasing for the next generation local wireless communication network.

Recently, the Chaotic Shape-forming Filter (CSF) was presented to encode information bits into chaotic signals \cite{Ren2019_Chaotic}. It has not only a corresponding matched filter to improve the Signal-to-Noise Ratio (SNR) at the receiver \cite{Corron2010_A}, which is not achievable by traditional chaotic signal generated by Logistic or Tent map even continuous chaotic signal generated by, for example, Lorenz system, but also it has ability to resist inter-symbol interference \cite{Yao2017_Chaos,Ren2013_Wireless}, by this way, it improves the performance of communications systems in complex radio channels \cite{Bai2020_Radio,Ren2020_Performance}. Moreover, the CSF can be implemented by using simple electronic circuit \cite{Ren2016_Experimental} or digital signal processor \cite{Yao2019_Experimental}, which addressed the demands of low consumption, low complexity and low power consumption in IoT communication. The chaotic signal generated by CSF can be used as spread sequence to achieve not only double stream data transmission \cite{Bai2019_Experimental} but also higher reliability as compared to the conventional chaotic spread sequence communications \cite{Bai2020_Double}. However, the current CSF researches only provide single chaotic carrier mostly for single user, which limits its application in multi-user cases.

In this work, inspired by the fundamental idea of the CSF, a Chaotic Pseudo Orthogonal Carriers Multi-Access (CPOCMA) communication system is proposed to provide multi-channel multiuser information transmission with improved spectrum efficiency. The pseudo orthogonal chaotic signals are generated by a Chaotic Pseudo Orthogonal Shape-forming Filter (CPOSF) bank at the transmitter which provides multiuser transmission in a time slot using single physical frequency carrier. At the receiver, the received signal is fed parallel into a Chaotic Pseudo Orthogonal Correlation Filter (CPOCF) bank and a Chaotic Pseudo Orthogonal Matched Filter (CPOMF) bank to separate the multi-channel information signal and to maximize the SNR, respectively. In single-user case, the CPOCMA scheme shows lower BER compared with its competitors due to the utilization of CPOMF. In multi-user case, the CPOCMA scheme shows higher throughput and lower BER with narrower bandwidth due to the use of CPOCF. The contributions of this work are as follows:

(i) A CPOCF bank is proposed to separate the pseudo orthogonal chaotic signals generated by the CPOSF bank, and achieve higher frequency spectrum utilization rate compared to the conventional multi access communication scheme such as Code Division Multiple Access (CDMA) and Frequency Division Multiple Access (FDMA).

(ii) A new CPOCMA modulation scheme offers multi-carrier multiuser information transmission with the higher spectrum efficiency as compared to conventional chaotic communications schemes.

(iii) An averaging operation is designed to decode part of parallel information bits transmitted by different chaotic carriers, then sort operation is used to decode the rest information bits, by this way, the better noise resistance and BER performance are achieved.

(iv) The analytical bit error rate expression is derived for CPOCMA in Additive White Gaussian Noise (AWGN) channel.

(v) Simulation and the experiment results based on Wireless open-Access Research Platform (WARP) are performed to show the superiority of the proposed scheme.

The rest of this paper is organized as follows: In Section II, the proposed CPOCMA system is described in detail. The performance of system is analyzed in Section III. Numerical simulations and discussions are presented in Section IV. The experiment validation and results analysis are given in Section V, and the conclusion remarks are presented in Section VI.

\section{CPOCMA System Architecture}
In this section we present the CPOCMA system design. The aim of the proposed system is to increase the BTR, to enhance the spectrum efficiency and to extend the CSF-based communications for multi-user application.
\subsection{The CPOCMA Transmitter}
\begin{figure}[!t]
\centering
\includegraphics[width=3.5in]{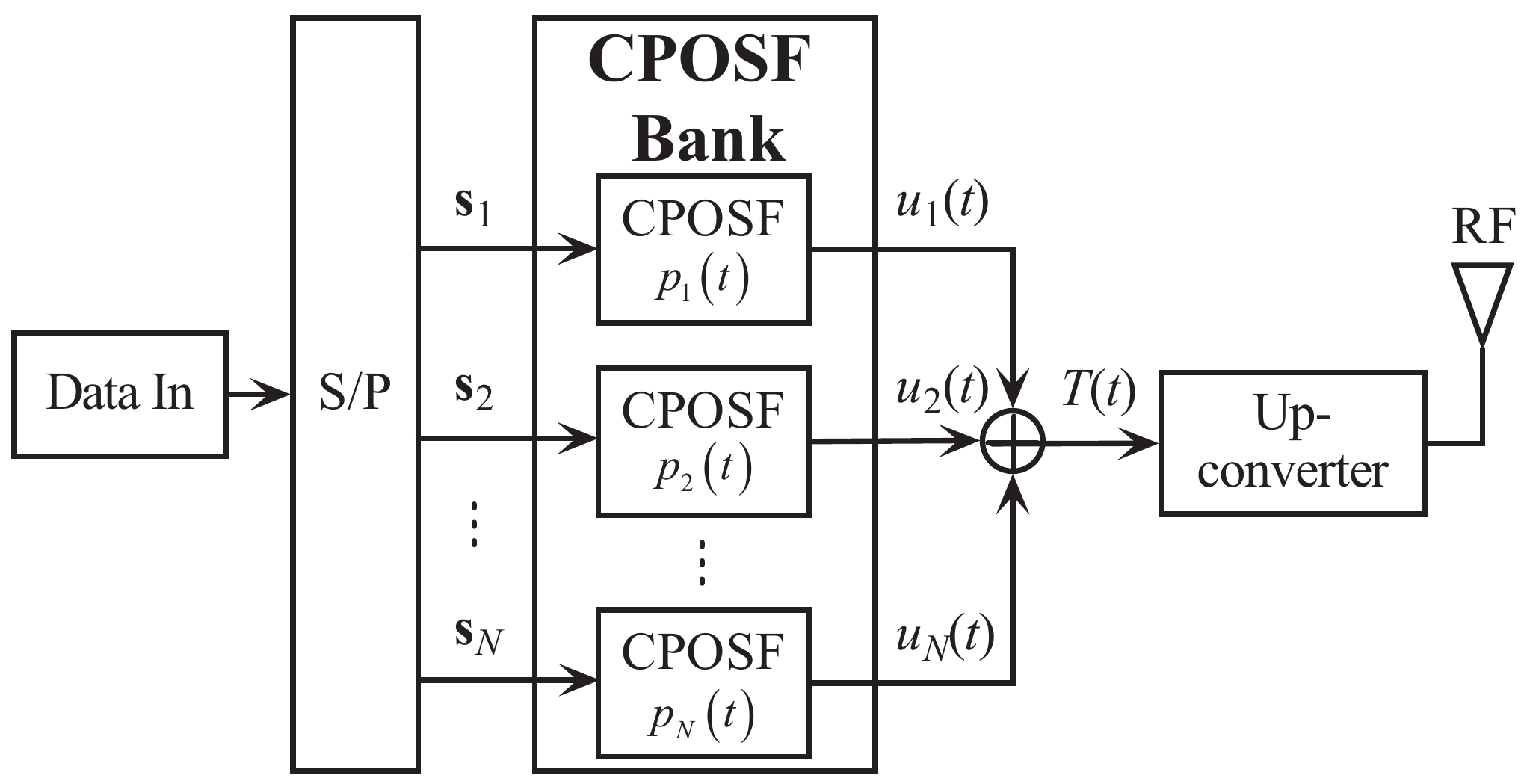}%Fig1.eps
\caption{Block diagram of the CPOCMA transmitter.}
\label{Fig_CPOCMA_transmitter}
\end{figure}

The block diagram of the CPOCMA transmitter is shown in Fig. \ref{Fig_CPOCMA_transmitter}. In this system, we consider $N$ chaotic subcarriers with the same central physical frequency to transmit $N$ information sequences simultaneously.

The input information sequence is first converted into $N$ parallel data sequences ${{\bf{s}}_n}$ for $n = {\rm{ 1}},{\rm{ 2}},{\rm{ }} \ldots ,N.$
\begin{equation}\label{Eq_sn}
{{\bf{s}}_n} = \left[ {{s_{n,1}},...{s_{n,m}},...,{s_{n,M}}} \right],
\end{equation}
where ${s_{n,m}} \in \left\{ { \pm 1} \right\}$ is the $m$th bit of the $n$th information sequence, $M$ is the number of the bits in each information sequence.

\begin{figure*}
\begin{minipage}[!t]{0.5\linewidth}
\centering
  \subfigure[ ]{
    \label{Fig_basis_function} %% label for first subfigure
    \includegraphics[scale=0.52,trim= 0 0 0 0]{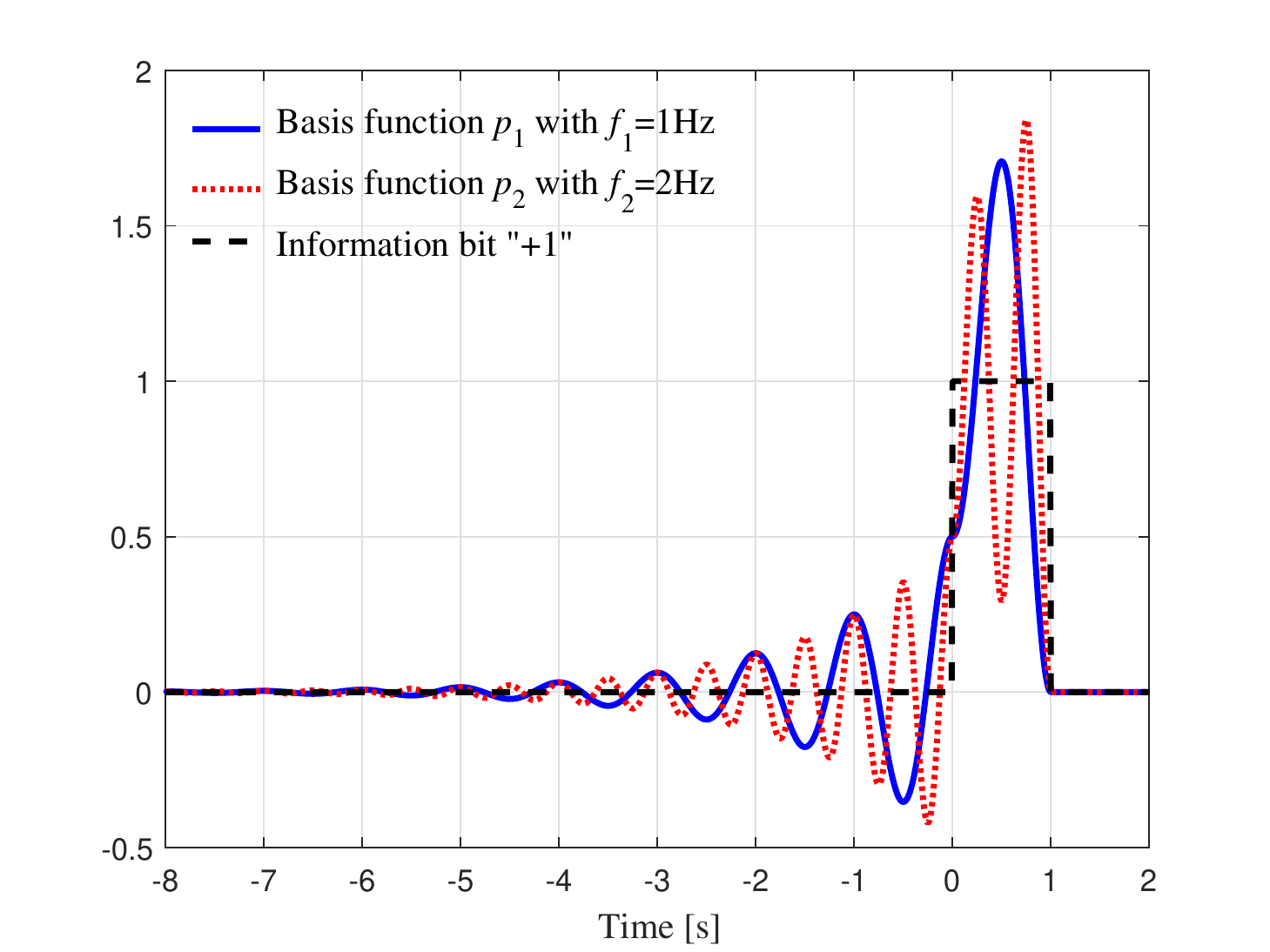}}
  \end{minipage}%
  \begin{minipage}[!t]{0.5\linewidth}
 \centering
  \subfigure[ ]{
    \label{Fig_chaotic_signals} %% label for second subfigure
    \includegraphics[scale=0.52,trim= 0 0 0 0]{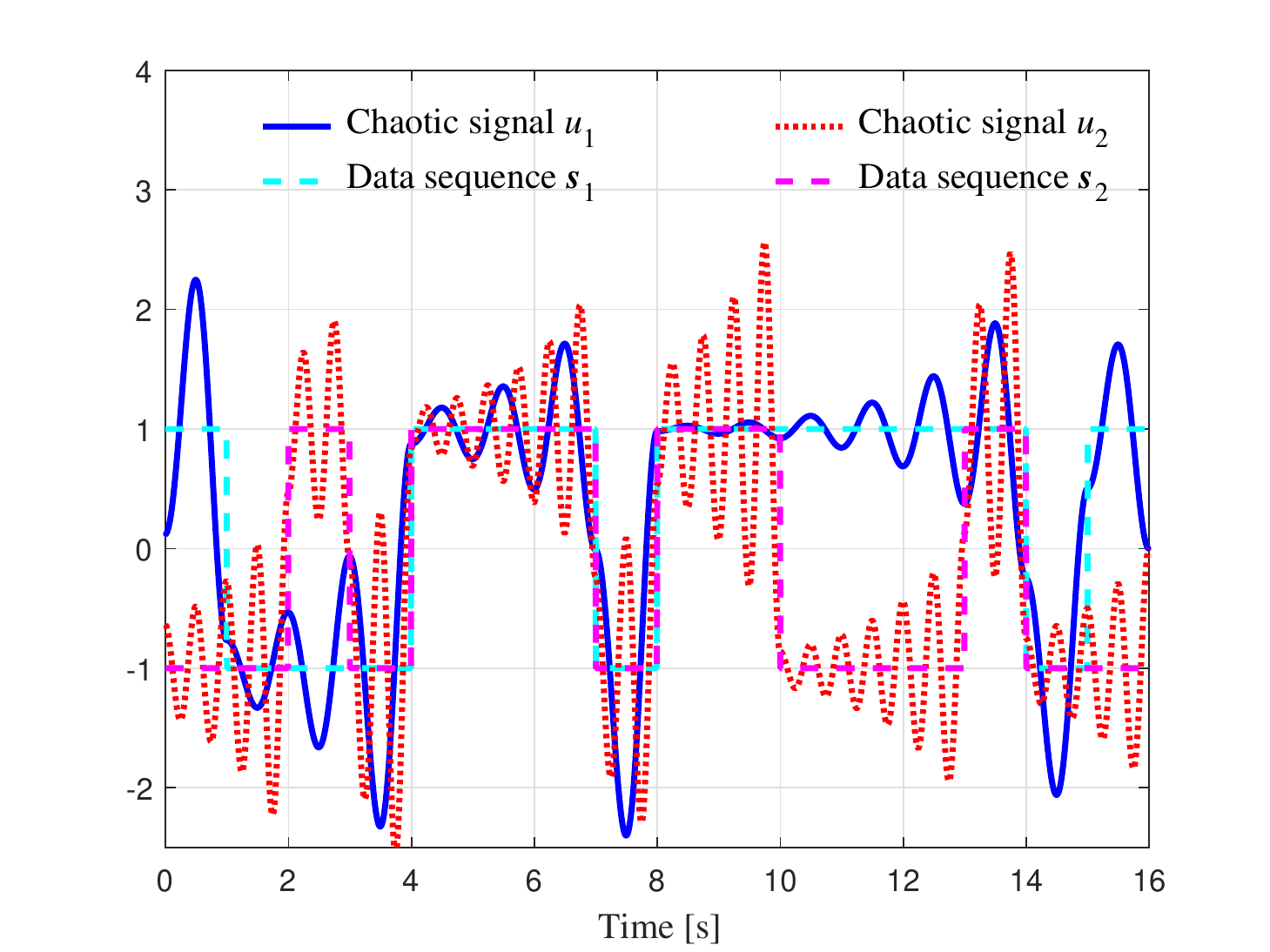}}
    \end{minipage}%
    \\
    \caption{The base function and the output example of the CPOSF bank. (a) The bases function curves and the information bit "${s_{{\rm{1}},{\rm{1}}}} = {s_{{\rm{2}},{\rm{1}}}} = {\rm{ }} + {\rm{1}}$" with the initial frequency $f$ = 1Hz, the base frequencies ${f_1}$ = 1 Hz and ${f_2}$ =2 Hz, respectively. (b) the chaotic signals generated by two data sequences using two base frequencies given in subplot (a) and their corresponding information bits.}
  \label{Fig_example_chaotic_signals} %% label for entire figure
\end{figure*}

After the Serial-to-Parallel (S/P) conversion, the information sequence ${{\bf{s}}_n}$ is fed into the corresponding CPOSF to generate output subcarrier ${u_n}\left( t \right)$ given by
\begin{equation}\label{Eq_CPOSF}
{u_n}\left( t \right) = \sum\limits_{m = \left\lfloor t \right\rfloor }^{\left\lfloor t \right\rfloor  + \infty } {{s_{n,m}}\cdot{p_n}\left( {t - m} \right)},
\end{equation}
where $\left\lfloor t \right\rfloor$ indicates the largest integer less than or equal to $t$, the $n$th basis function ${p_n}\left( t \right)$ in CPOSF bank is given by
\begin{small}
\begin{equation}\label{Eq_basis_functions}
\begin{array}{l}
 {p_n}\left( t \right) =  \\
 \left\{ {\begin{array}{*{20}{c}}
   {\left( {1 - {e^{ - \frac{\beta }{f}}}} \right){e^{\beta t}}\left( {\cos \left( {{\omega _n}t} \right) - \frac{\beta }{{{\omega _n}}}\sin \left( {{\omega _n}t} \right)} \right),} & {t < 0}  \\
   {1 - {e^{\beta \left( {t - \frac{1}{f}} \right)}}\left( {\cos \left( {{\omega _n}t} \right) - \frac{\beta }{{{\omega _n}}}\sin \left( {{\omega _n}t} \right)} \right),} & {0 \le t < \frac{1}{f}}  \\
   {0,} & {t \ge \frac{1}{f}},  \\
\end{array}} \right. \\
 \end{array}
\end{equation}\end{small}
where the system parameters $\beta  = f\ln 2$, ${\omega _n} = 2\pi {f_n}$, $f$ is the initial frequency and ${f_n}$ is the base frequency of the $n$th basis function. Using two base frequencies as an example, the curves of two bases functions ${p_1}$ (solid line) and ${p_2}$ (dotted line), and the same information bit "${s_{{\rm{1}},{\rm{1}}}} = {s_{{\rm{2}},{\rm{1}}}} = {\rm{ }} + {\rm{1}}$" (dashed line) are given in Fig. \ref{Fig_basis_function} with the initial frequency $f$ = 1 Hz, and the base frequencies ${f_1}$ = 1 Hz and ${f_2}$ = 2 Hz, respectively. Figure \ref{Fig_chaotic_signals} shows chaotic signals ${u_{\rm{1}}}\left( t \right)$ and ${u_{\rm{2}}}\left( t \right)$ generated by the data sequences
${{\bf{s}}_1}$ = [1, -1, -1, -1, 1, 1, 1, -1, 1, 1, 1, 1, 1, 1, -1, 1]
and
${{\bf{s}}_2}$ = [-1, -1, 1, -1, 1, 1, 1, -1, 1, 1, -1, -1, -1, 1, -1, -1]
and their corresponding symbols, respectively, where the initial frequency, base frequencies and $\beta$ is the same as these in Fig. \ref{Fig_basis_function}.

The transmitted signal of the CPOCMA is the sum of all CPOSF outputs as given by
\begin{equation}\label{Eq_transmitted_signal}
T\left( t \right) = \sum\limits_{n = 1}^N {{u_n}\left( t \right)},
\end{equation}
where the modulated subcarriers ${u_n}\left( t \right)$ for $n = 1,...,N$ are pseudo-orthogonal chaotic signals. It means that an individual subcarrier can be extracted under the multiple access interference by a CPOCF bank in receiver, because the correlation function with others CPOCF is close to zero. The transmitted signal is sent to the wireless channel through the up-converter with single physical frequency and corresponding Radio Frequency (RF) antenna.

\subsection{The CPOCMA Receiver}
\begin{figure*}[!t]
\centering
\includegraphics[width=5in]{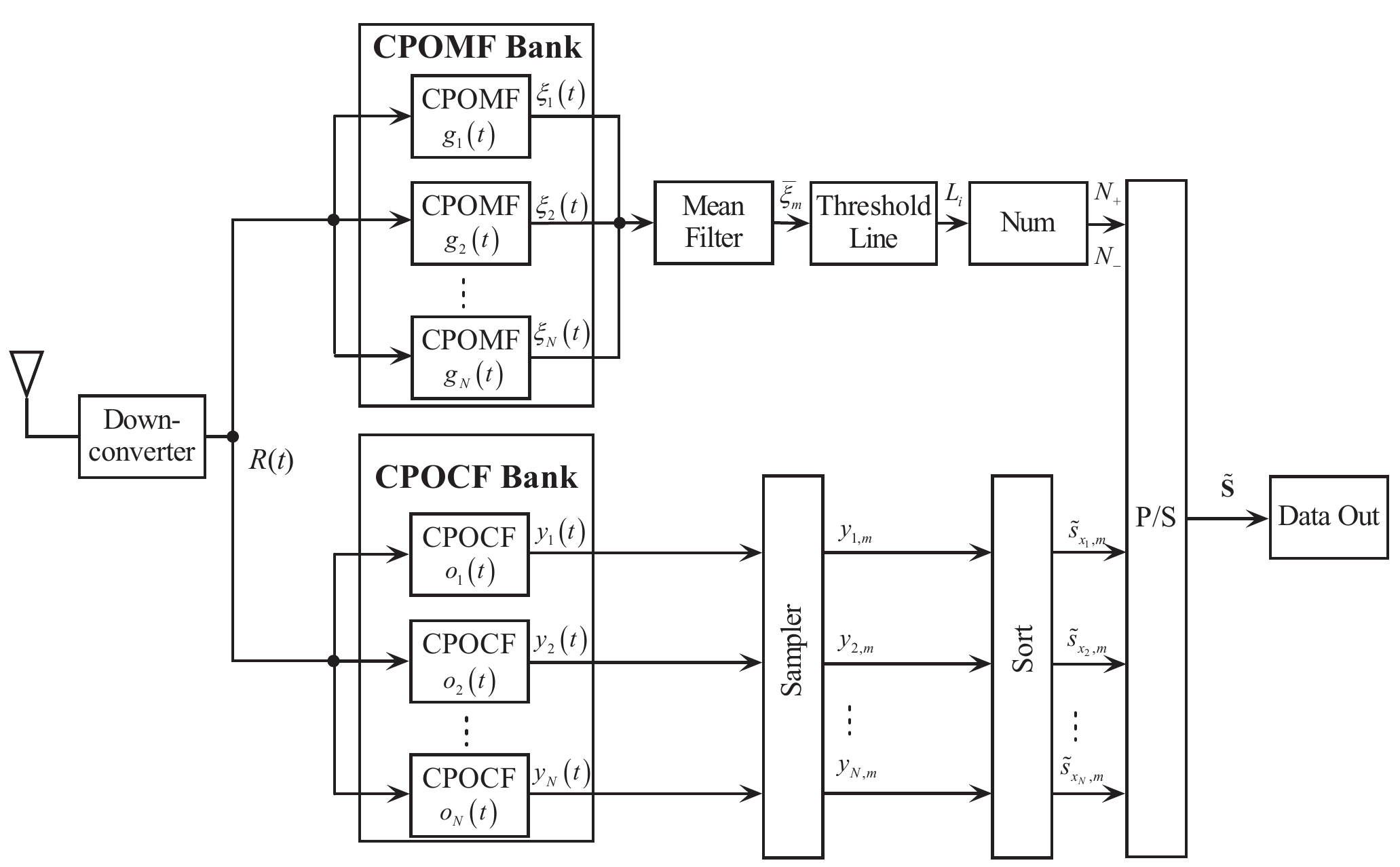}
\caption{Block diagram of the CPOCMA receiver.}
\label{Fig_CPOCMA_receiver}
\end{figure*}

The block diagram of the CPOCMA receiver is shown in Fig. \ref{Fig_CPOCMA_receiver}. After the transmitted signal passes through the wireless channel, the received signal is captured by the receiver antenna, which is passed through down-carrier as shown in Fig. \ref{Fig_CPOCMA_receiver} to remove the physical transmission frequency so as to obtain $R\left( t \right)$. Then, the received signal is fed into the CPOMF bank and the CPOCF bank, separately.

\begin{figure}
\begin{minipage}[!t]{1\columnwidth}
\centering
  \subfigure[ ]{
    \label{Fig_MF_signal} %% label for first subfigure
    \includegraphics[scale=0.55,trim= 0 0 0 0]{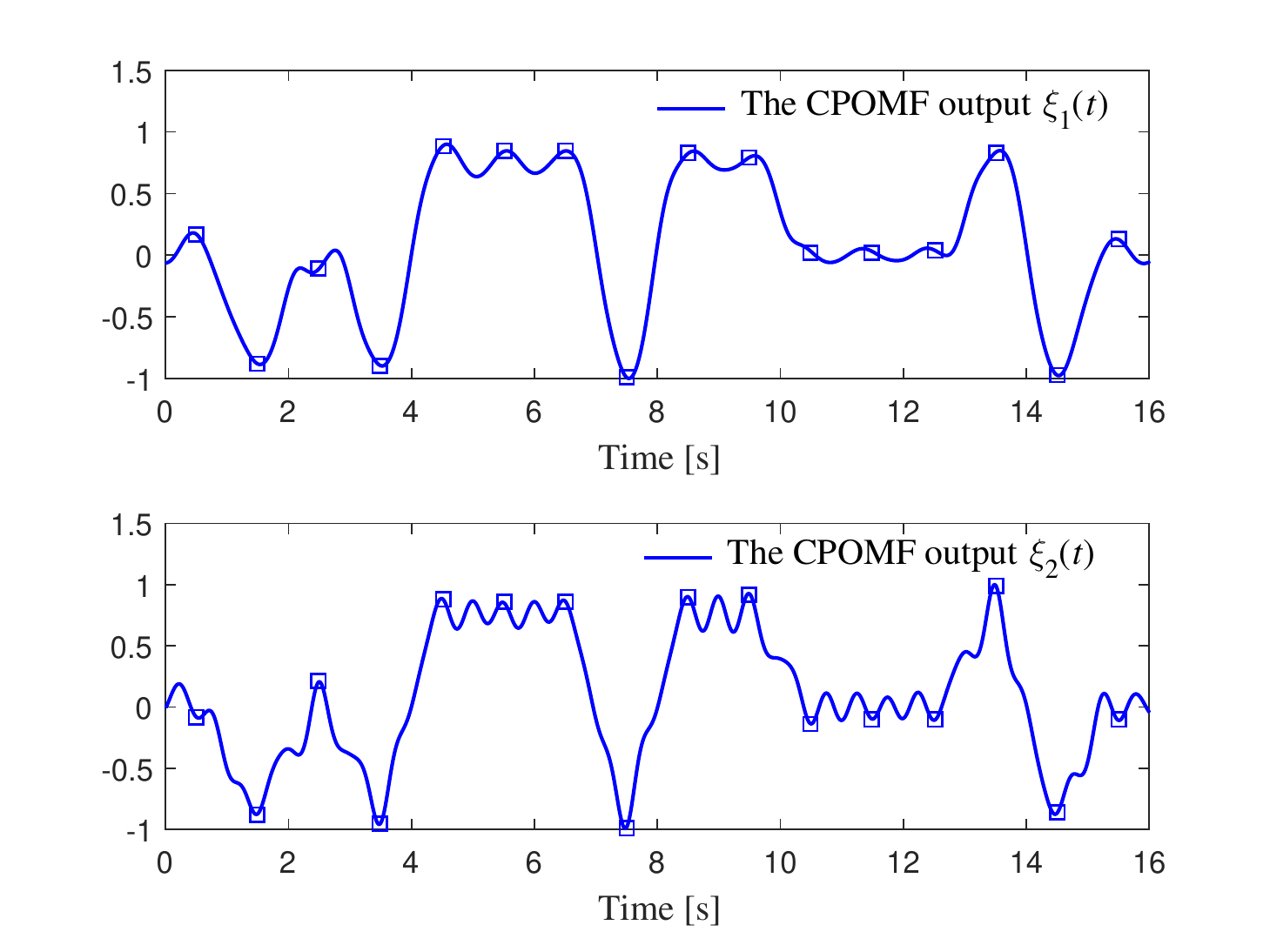}}
  \end{minipage}%
  \\
  \begin{minipage}[!t]{1\columnwidth}
 \centering
  \subfigure[ ]{
    \label{Fig_decision_line} %% label for second subfigure
    \includegraphics[scale=0.55,trim= 0 0 0 0]{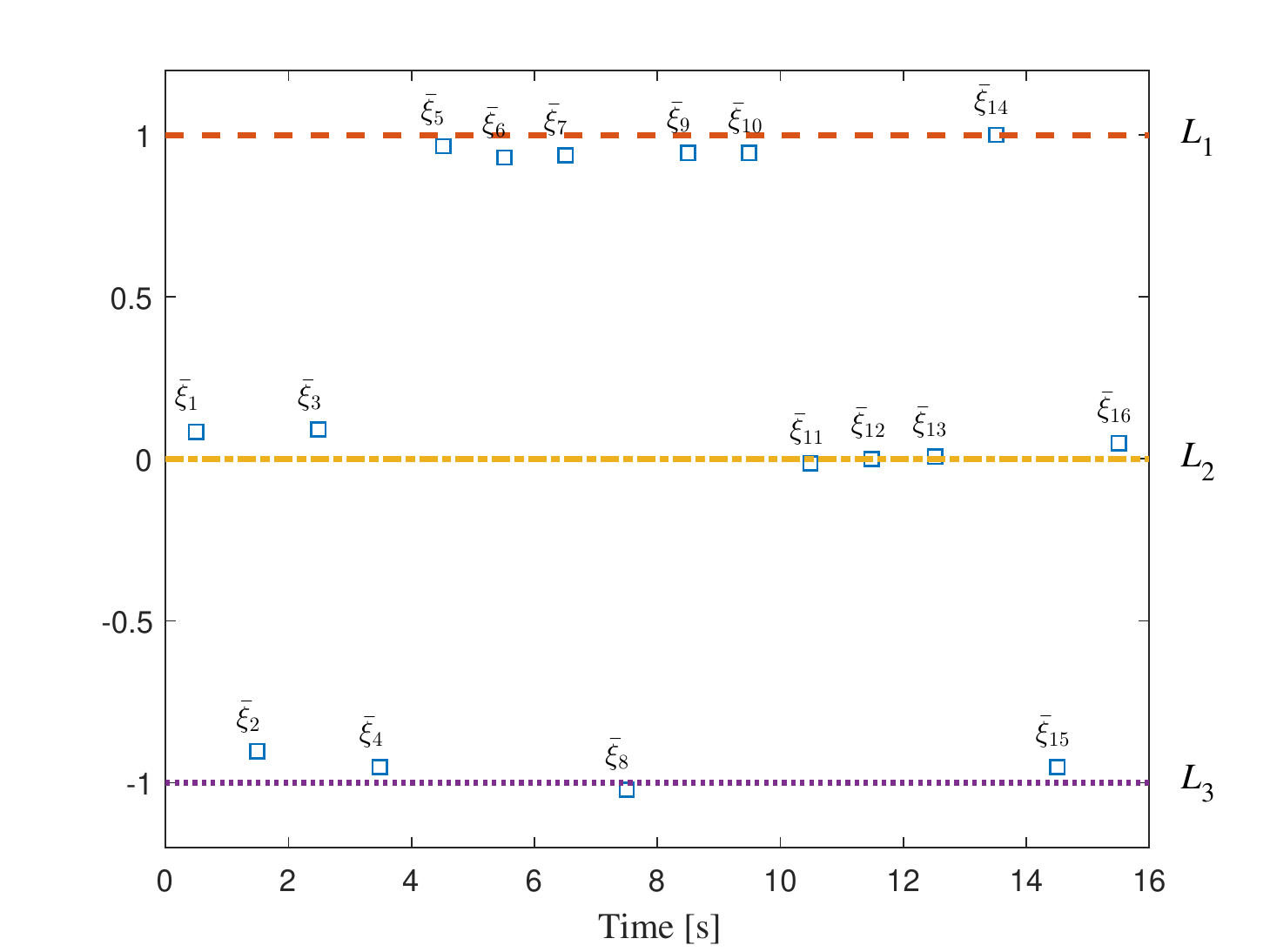}}
    \end{minipage}%
\\
  \begin{minipage}[!t]{1\columnwidth}
  \centering
  \subfigure[ ]{
    \label{Fig_CPOCF} %% label for second subfigure
    \includegraphics[scale=0.55,trim= 0 0 0 0]{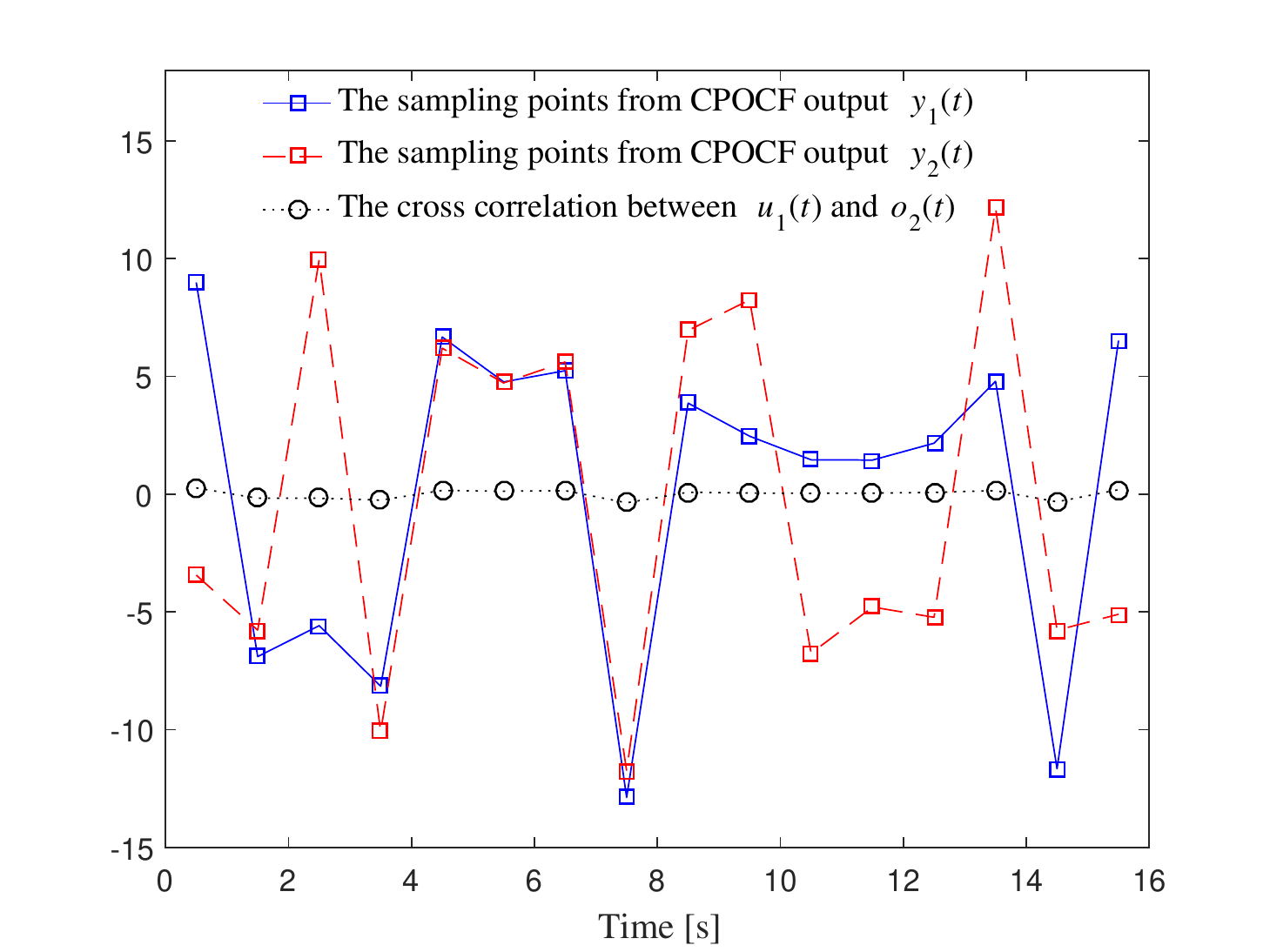}}
    \end{minipage}%
    \\
    \caption{The schematic illustration at the receiver. (a) The matched filter outputs ${\xi _1}\left( t \right)$ and ${\xi _2}\left( t \right)$ in the upper panel and in the lower panel, respectively, and the sampling points with maximum SNR by blue square marks; (b) the mean filter outputs and the corresponding decision line for $N$ = 2 subcarriers; (c) the chaotic pseudo-orthogonal filter bank outputs.}
  \label{Fig_example_rece} %% label for entire figure
\end{figure}

In the Chaotic Pseudo Orthogonal Matched Filter (CPOMF) bank, the $n$th output ${\xi _n}\left( t \right)$ is obtained from the corresponding $n$th CPOMF, as given by
\begin{equation}\label{Eq_CPOMA}
{\xi _n}\left( t \right) = \int\limits_{ - \infty }^\infty  {R\left( \tau  \right){g_n}\left( {t - \tau } \right)} d\tau ,
\end{equation}
where ${g_n}\left( t \right) = {p_n}\left( { - t} \right)$ is the time reverse of the $n$th basis function of the CPOCMA. The filter outputs corresponding to the signals shown in Fig. \ref{Fig_chaotic_signals} are given in the upper panel and the lower panel of Fig. \ref{Fig_MF_signal}. The sampling points (with maximum signal to noise ratio \cite{Corron2015_Chaos}) from the filter outputs using sampling interval ${T_c} = {\rm{1}}/f$ are marked in Fig. \ref{Fig_MF_signal} by blue squares, as given by
\begin{equation}\label{Eq_sum_points}
\begin{array}{*{20}{c}}
   {{\xi _{n,m}} = {\xi _n}\left( {{T_c}\left( {m - 1} \right) + {{{T_c}} \mathord{\left/
 {\vphantom {{{T_c}} 2}} \right.
 \kern-\nulldelimiterspace} 2}} \right),} & {1 \le m \le M}  \\
\end{array}.
\end{equation}
An averaging operation is performed to further reduce the influence of ISI and noise, as given by
\begin{equation}\label{Eq_averaging}
{\bf{\bar \xi }}{\rm{ = }}\left[ {{{\bar \xi }_1},{{\bar \xi }_2},...,{{\bar \xi }_m},...,{{\bar \xi }_M}} \right],
\end{equation}
where ${\bar \xi _m} = \frac{1}{N}\sum\limits_{n = 1}^N {{\xi _{n,m}}} ,$ $\left( {1 \le m \le M} \right)$,  $N$ is the number of data sequences. The averaging outputs are shown in Fig. \ref{Fig_decision_line} by blue square marks.

We define $N$+1 decision lines for the averaging operation output as
\begin{equation}\label{Eq_decision_line}
\begin{array}{*{20}{c}}
   {{L_i} = \left( {1 - \frac{2}{N}\left( {i - 1} \right)} \right)\Delta ,} & {1 \le i \le N + 1}  \\
\end{array},
\end{equation}
where $\Delta  = \max \left( {\left| {{\bf{\bar \xi }}} \right|} \right)$. Figure \ref{Fig_decision_line} shows three decision threshold lines plotted by red dashed line, yellow dash-dotted line and purple dotted line, respectively, corresponding to the signal transmitted in Fig. \ref{Fig_chaotic_signals}.
The index of the minimum distance between the $m$th mean filter output with respect to all $N$+1 decision lines is
\begin{equation}\label{Eq_decision_line1}
{D_m} = \mathop {\arg }\limits_{i \in \left[ {1,N + 1} \right]} \min \left\{ {{d_i}} \right\},
\end{equation}
where ${d_i} = \sqrt {{{\left( {{L_i} - {{\bar \xi }_m}} \right)}^2}} ,\left( {1 \le i \le N + 1} \right)$. It means that there are ${N_{\rm{ + }}} = N - \left( {{D_m} - 1} \right)$ information bits being "+1" and ${N_ - } = {D_m} - 1$ information bits being "$-$1" in the time slot $m$, i.e.,
\begin{equation}\label{Eq_Num_1}
{\tilde s_{1 \to N,m}} = {\left[ {\underbrace { + 1,..., + 1}_{{N_ + } = N - \left( {{D_m} - 1} \right)},\underbrace { - 1,..., - 1}_{{N_ - } = {D_m} - 1}} \right]^T},
\end{equation}
where ${\tilde s_{1 \to N,m}}$ represents the recovered $N$ information bits with ${N_ + }$ "1"s and ${N_ - }$ "$-$1"s according to $m$th averaging output. In the illustrated example of Fig. \ref{Fig_decision_line}, ${D_{\rm{1}}} = 2\left( {{N_{\rm{ + }}}{\rm{ = }}1,{N_ - }{\rm{ = }}1} \right)$ means that one information bit is "+1" and one information bit is "$-$1" in the first time slot, but we cannot confirm whether the recovered bits are (${\tilde s_{1,1}}{\rm{ = }}1$, ${\tilde s_{2,1}} =  - 1$) or (${\tilde s_{1,1}} =  - 1$, ${\tilde s_{2,1}} = 1$). However, ${D_{\rm{2}}} = 3\left( {{N_{\rm{ + }}}{\rm{ = 0}},{N_ - }{\rm{ = 2}}} \right)$ means that two information bits "$-$1" in this time slot, i.e., the recovered bits are ${\tilde s_{1,2}}{\rm{ = }}{\tilde s_{2,2}}{\rm{ = }} - 1$. From above illustration, we know that we cannot ensure to decode the information only from the output of the averaging operation except some extreme cases, although we can determine the number of information "$+$1" and "$-$1" in the time slot. Therefore, we need the help of the CPOCF given in the following to decode the information bits.

The output ${y_n}\left( t \right)$ of the $n$-th Chaotic Pseudo Orthogonal Correlation Filter (CPOCF) is obtained by
\begin{equation}\label{Eq_CPOCF}
{y_n}\left( t \right) = \int\limits_{ - \infty }^\infty  {R\left( \tau  \right){o_n}\left( {t - \tau } \right)} d\tau ,
\end{equation}
where the orthogonal basis function ${o_n}\left( t \right)$ is given by
\begin{small}
\begin{equation}\label{Eq_orth_basis_function}
\begin{array}{l}
 {o_n}\left( t \right) =  \\
 \left\{ {\begin{array}{*{20}{c}}
   {\left( {1 - {e^{ - \beta }}} \right){e^{\beta t}}\left( {\cos \left( {{\omega _n}t} \right) - \frac{\beta }{{{\omega _n}}}\sin \left( {{\omega _n}t} \right)} \right)} & {t < 0}  \\
   { - {e^{\beta \left( {t - 1} \right)}}\left( {\cos \left( {{\omega _n}t} \right) - \frac{\beta }{{{\omega _n}}}\sin \left( {{\omega _n}t} \right)} \right)} & {0 \le t < \frac{1}{f}}  \\
   0 & {t \ge \frac{1}{f}}.  \\
\end{array}} \right. \\
 \end{array}
\end{equation}\end{small}
The sampling points from ${y_n}\left( t \right)$ with sampling interval ${T_c} = {\rm{1}}/f$ are denoted as
\begin{equation}\label{Eq_CF_SUM}
\begin{array}{*{20}{c}}
   {{y_{n,m}} = {y_n}\left( {{T_c}\left( {m - 1} \right) + {{{T_c}} \mathord{\left/
 {\vphantom {{{T_c}} 2}} \right.
 \kern-\nulldelimiterspace} 2}} \right),} & {1 \le m \le M}  \\
\end{array}.
\end{equation}
In the illustrated example given in Fig. \ref{Fig_chaotic_signals}, ${y_{\rm{1}}}\left( t \right)$ and ${y_{\rm{2}}}\left( t \right)$ are shown by blue solid line with square marks for the sampling point ${y_{{\rm{1}},m}}$ and red dashed line with square marks for ${y_{{\rm{2}},m}}$, respectively, in Fig. \ref{Fig_CPOCF}. In order to verify the orthogonality of different subcarriers, the cross correlation between subcarrier ${u_{\rm{1}}}\left( t \right)$ and orthogonal basis function ${o_{\rm{2}}}\left( t \right)$ is shown in Fig. \ref{Fig_CPOCF} by black dotted line with circle marks. It can be seen from the black dotted line with circle marks in Fig. \ref{Fig_CPOCF} that the magnitude of the cross correlation is near zero, and far less than the corresponding orthogonal filter outputs, i.e., the square marks in Fig. \ref{Fig_CPOCF}. Then, a descending order sorting operation is performed for $\left[ {{y_{{\rm{1}},m}},{y_{{\rm{2}},m}}, \ldots ,{y_N}_{,m}} \right]$ in $m$th time slot. $\left[ {{x_1},{x_2}, \ldots ,{x_N}} \right]$ denotes the indices after sorting operation, which means that ${y_{{x_1},m}} > {y_{{x_2},m}} > ... > {y_{{x_N},m}}$. By dividing the sorted sequence into to group according to ${N_ + }$ and ${N_ - }$, we can decode $N$ transmitted information bits at the $m$th time slot as
\begin{align*}\label{Eq_recover}
\begin{array}{l}
 {\left[ {{{\tilde s}_{{x_1},m}},{{\tilde s}_{{x_2},m}},...,{{\tilde s}_{{x_N},m}}} \right]^T} = {{\tilde s}_{1 \to N,m}} \\
  = {\left[ {\underbrace { + 1,..., + 1}_{{N_ + } = N - \left( {{D_m} - 1} \right)},\underbrace { - 1,..., - 1}_{{N_ - } = {D_m} - 1}} \right]^T}, \\
 \end{array}
\end{align*}
where ${\tilde s_{1 \to N,m}}$ is defined in Eq. (\ref{Eq_Num_1}).

In the illustrated example, when $m$ = 1, ${y_{{\rm{1}},{\rm{1}}}} > {y_{{\rm{2}},{\rm{1}}}}$ means that the indices are ${x_1} = 1$ and ${x_2} = 2$. Combined with the number of information bits "+1" and "$-$1" are ${N_{\rm{ + }}}{\rm{ = }}1$, ${N_{ - 1}} = 1$ in Eq. (\ref{Eq_Num_1}), the information bits for the $m$ = 1 sampling instant are recovered as $\left[ {{{\tilde s}_{1,1}},{{\tilde s}_{2,1}}} \right]{\rm{ = }}\left[ { + 1, - 1} \right]$. For $m$ = 2 sampling instant, the recovered information bits are $\left[ {{{\tilde s}_{1,2}},{{\tilde s}_{2,2}}} \right]{\rm{ = }}\left[ { - 1, - 1} \right]$, which can be directly decoded according to Eq. (\ref{Eq_Num_1}) without the help of the CPOMF. For the $m$ = 3 sampling instant, ${y_{{\rm{1}},{\rm{3}}}} < {y_{{\rm{2}},{\rm{3}}}}$ $\left( {{x_1} = 2,{x_2} = 1} \right)$ means that the recovered order is $\left[ {{{\tilde s}_{2,3}},{{\tilde s}_{1,3}}} \right]$, and the recovered information are $\left[ {{{\tilde s}_{1,3}},{{\tilde s}_{2,3}}} \right]{\rm{ = }}\left[ { - 1, + 1} \right]$ according to the result of Eq. (\ref{Eq_Num_1}). To this end, all information bits can be decoded completely and sequentially.
%\begin{figure*}
%\begin{minipage}[!t]{0.5\linewidth}
%\centering
%  \subfigure[ ]{
%    \label{Fig_MF_signal} %% label for first subfigure
%    \includegraphics[scale=0.5,trim= 0 0 0 0]{Fig4a.eps}}
%  \end{minipage}%
%  \begin{minipage}[!t]{0.5\linewidth}
% \centering
%  \subfigure[ ]{
%    \label{Fig_decision_line} %% label for second subfigure
%    \includegraphics[scale=0.5,trim= 0 0 0 0]{Fig4b.eps}}
%    \end{minipage}%
%
%  \begin{minipage}[!t]{1\columnwidth}
%  \centering
%  \subfigure[ ]{
%    \label{Fig_CPOCF} %% label for second subfigure
%    \includegraphics[scale=0.5,trim= 0 0 0 0]{Fig4c.eps}}
%    \end{minipage}%
%    \\
%    \caption{The schematic illustration at the receiver. (a) The matched filter outputs ${\xi _1}\left( t \right)$ and ${\xi _2}\left( t \right)$ in the upper panel and in the lower panel, respectively, and the sampling points with maximum SNR by blue square marks; (b) the mean filter outputs and the corresponding decision line for $N$ = 2 subcarriers; (c) the chaotic pseudo-orthogonal filter bank outputs.}
%  \label{Fig_example_rece} %% label for entire figure
%\end{figure*}

\section{Performance Analysis of CPOCMA System}
In this section, the performance of the CPOCMA system is evaluated and the BER expression is derived. In order to simplify the analysis, the case of $N$ = 2 subcarriers is considered here, and other cases can be calculated by the same way.

The received signal with AWGN is given by
\begin{equation}\label{Eq_add_noise}
R\left( t \right) = \sum\limits_{n = 1}^{N{\rm{ = }}2} {{u_n}\left( t \right)} {\rm{ + }}w\left( t \right),
\end{equation}
where ${u_n}\left( t \right)$ is the $n$th subcarrier in the transmitted signal and $w\left( t \right)$ denotes the additive white Gaussian noise. The received signal is sent into the CPOMF bank, and its outputs are
\begin{equation}\label{Eq_MF_filter_out1}
\begin{array}{l}
 {\xi _1}\left( t \right) = \int\limits_{ - \infty }^\infty  {R\left( \tau  \right){g_1}\left( {t - \tau } \right)} d\tau  \\
 {\rm{ = }}\sum\limits_{m = 1}^M {{s_{1,m}}} \int\limits_{ - \infty }^\infty  {{p_1}\left( \tau  \right){p_1}\left( {\tau  - t + m/f} \right)} d\tau  \\
  + \sum\limits_{m = 1}^M {{s_{2,m}}} \int\limits_{ - \infty }^\infty  {{p_2}\left( \tau  \right){p_1}\left( {\tau  - t + m/f} \right)} d\tau  \\
  + \int\limits_{ - \infty }^\infty  {w\left( \tau  \right){p_1}\left( {\tau  - t} \right)} d\tau  \\
 \end{array}
\end{equation}
and
\begin{equation}\label{Eq_MF_filter_out2}
\begin{array}{l}
 {\xi _2}\left( t \right) = \int\limits_{ - \infty }^\infty  {R\left( \tau  \right){g_2}\left( {t - \tau } \right)} d\tau  \\
 {\rm{ = }}\sum\limits_{m = 1}^M {{s_{2,m}}} \int\limits_{ - \infty }^\infty  {{p_2}\left( \tau  \right){p_2}\left( {\tau  - t + m/f} \right)} d\tau  \\
  + \sum\limits_{m = 1}^M {{s_{1,m}}} \int\limits_{ - \infty }^\infty  {{p_1}\left( \tau  \right){p_2}\left( {\tau  - t + m/f} \right)} d\tau  \\
  + \int\limits_{ - \infty }^\infty  {w\left( \tau  \right){p_2}\left( {\tau  - t} \right)} d\tau,  \\
 \end{array}
\end{equation}
where $m\left( {m = {\rm{1}}, \ldots ,M} \right)$ denotes the $m$th bit in the $n$th $( {n = {\rm{1}},{\rm{2}}}$ here$)$ subcarrier. Without loss of generality, the $j$th $\left( {j = {\rm{1}}, \ldots ,M} \right)$ sampling points ${\xi _{1,j}}$ and ${\xi _{2,j}}$ from ${\xi _1}\left( t \right)$ and ${\xi _2}\left( t \right)$ are given as
\begin{equation}\label{Eq_MF_sampling1}
{\xi _{1,j}}{\rm{ = }}{s_{1,j}}{E_1} + \sum\limits_{\scriptstyle m = 1 \hfill \atop
  \scriptstyle m \ne j \hfill}^M {{s_{1,m}}{I_{1,m}}}  + {s_{2,j}}{B_{1,j}} + {\Phi _{1,m}} + {W_{{M_1}}}
\end{equation}
and
\begin{equation}\label{Eq_MF_sampling2}
{\xi _{2,j}}{\rm{ = }}{s_{2,j}}{E_2} + \sum\limits_{\scriptstyle m = 1 \hfill \atop
  \scriptstyle m \ne j \hfill}^M {{s_{2,m}}{I_{2,m}}}  + {s_{1,j}}{B_{2,j}} + {\Phi _{2,m}} + {W_{{M_2}}},
\end{equation}
where the first term in Eqs. (\ref{Eq_MF_sampling1}-\ref{Eq_MF_sampling2}) is the expected energy and equal to the autocorrelation of the basis function, given by
\begin{equation}\label{Eq_MF_pro1}
\begin{array}{*{20}{c}}
   {{E_n}{\rm{ = 1 + }}\frac{{\left( {1 - {e^{ - \beta }}} \right)}}{2}\left[ {{\Delta _{n,1}} - \frac{{6\beta }}{{{\beta ^2} + \omega _n^2}}} \right],} & {n = 1,2}  \\
\end{array},
\end{equation}
where ${\Delta _{n,1}}{\rm{ = }}\left( {1 + \frac{{{\beta ^2}}}{{\omega _n^2}}} \right)\frac{1}{\beta } + \left( {1 - \frac{{{\beta ^2}}}{{\omega _n^2}}} \right)\frac{\beta }{{{\beta ^2} + \omega _n^2}}$.

The second term is the ISI caused by the subcarrier itself and
\begin{equation}\label{Eq_MF_pro2}
\begin{array}{*{20}{c}}
   \begin{array}{l}
 {I_{n,m}}{\rm{ = }}\frac{1}{4}\left( {{\Delta _{n,1}} + \frac{{2\beta }}{{{\beta ^2} + \omega _n^2}}} \right)\left( {1 - \frac{{8\beta }}{{{\beta ^2} + \omega _n^2}}} \right) \\
 \left( {1 - {e^{ - \beta }}} \right)\left( {{e^{ - \beta m}} - {e^{\beta \left( {1 - m} \right)}}} \right), \\
 \end{array} & {n = 1,2}.  \\
\end{array}
\end{equation}

The third term is the ISI caused by the $j$th sampling instant of other subcarriers and
\begin{equation}\label{Eq_MF_pro3}
\begin{array}{l}
 {B_{n,j}} =  \\
 \left( {1 - {e^{ - \beta }}} \right)\left( {{\Delta _{n,2}} - 2\beta \frac{{2{\beta ^2} + \sum\limits_{n = 1}^{N = 2} {\omega _n^2} }}{{{\beta ^4}{\rm{ + }}{\beta ^2}\sum\limits_{n = 1}^{N = 2} {\omega _n^2} {\rm{ + }}\prod\limits_{n = 1}^{N = 2} {\omega _n^2} }}} \right) + 1, \\
 \end{array}
\end{equation}
where ${\Delta _{n,2}}$ is given as
\begin{equation}\label{Eq_MF_pro4}
\begin{array}{l}
 {\Delta _{n,2}}{\rm{ = }}\left( {1 - \frac{{{\beta ^2}}}{{{\omega _1}{\omega _2}}}} \right)\frac{{2\beta }}{{4{\beta ^2} + {{\left( {{\omega _1}{\rm{ + }}{\omega _2}} \right)}^2}}} + \left( {1 - \frac{{{\beta ^2}}}{{{\omega _1}{\omega _2}}}} \right)\frac{{2\beta }}{{4{\beta ^2} + {\omega ^2}}} \\
 {\rm{ + }}\left( {\frac{\beta }{{{\omega _1}}}{\rm{ + }}\frac{\beta }{{{\omega _2}}}} \right)\frac{{{\omega _1}{\rm{ + }}{\omega _2}}}{{4{\beta ^2}{\rm{ + }}{{\left( {{\omega _1}{\rm{ + }}{\omega _2}} \right)}^2}}}{\rm{ + }}\left| {\frac{\beta }{{{\omega _1}}} - \frac{\beta }{{{\omega _2}}}} \right|\frac{\omega }{{4{\beta ^2} + {\omega ^2}}} \\
 \end{array},
\end{equation}
and $\omega {\rm{ = }}\left\{ {\begin{array}{*{20}{c}}
   {{\omega _1} - {\omega _2},} & {n = 1}  \\
   {{\omega _2} - {\omega _1},} & {n = 2}  \\
\end{array}} \right.$ in Eq. (\ref{Eq_MF_pro4}).

The fourth term is the ISI of the $m$th ($m = 1,...,M$ and $m \ne j$) symbol in other subcarriers to the $j$th sampling instant, which can be ignored due to ${\Phi _{n,m}} \ll {B_{n,j}}$.

The fifth term ${W_{{M_n}}}$ is filtered noise.

The $j$th mean filter output is given by
\begin{equation}\label{Eq_MF_pro5}
{{\bar \xi }_j} = \frac{1}{2}\sum\limits_{n = 1}^{N{\rm{ = }}2} {{\xi _{n,j}}} {\rm{ = }}\frac{1}{2}\left( \begin{array}{l}
 \sum\limits_{n = 1}^{N = 2} {{s_{n,j}}{E_n}}  + \sum\limits_{n = 1}^{N = 2} {{s_{n,j}}{B_{n,j}}}  \\
  + \sum\limits_{n = 1}^{N = 2} {\sum\limits_{\scriptstyle m = 1 \hfill \atop
  \scriptstyle m \ne j \hfill}^M {{s_{n,m}}{I_{n,m}}} }  + \sum\limits_{n = 1}^{N = 2} {{W_n}}  \\
 \end{array} \right).
\end{equation}
When subcarrier $N = {\rm{2}}$ and the transmitted signal is $\left( {{s_{1,j}},{s_{2,j}}} \right) = \left( { + 1, + 1} \right)$ or $\left( {{s_{1,j}},{s_{2,j}}} \right) = \left( { - 1, - 1} \right)$, a judgment threshold can be obtained
\begin{equation}\label{Eq_MF_pro6}
\Delta {\rm{ = }}\left| {{{\bar \xi }_j}} \right|.
\end{equation}
It means that the number ${N_{\rm{ + }}}$ and ${N_ - }$ for arbitrary information bits "$+$1" and "$-$1" at $j$th sampling instant can be obtained by MF, given as
\begin{equation}\label{Eq_MF_pro7}
\left\{ {\begin{array}{*{20}{c}}
   {{N_{\rm{ + }}}{\rm{ = }}\left\{ {\begin{array}{*{20}{c}}
   {2,} & {{{\bar \xi }_j} > {\Delta  \mathord{\left/
 {\vphantom {\Delta  2}} \right.
 \kern-\nulldelimiterspace} 2}}  \\
   {1,} & { - {\Delta  \mathord{\left/
 {\vphantom {\Delta  2}} \right.
 \kern-\nulldelimiterspace} 2} < {{\bar \xi }_j} \le {\Delta  \mathord{\left/
 {\vphantom {\Delta  2}} \right.
 \kern-\nulldelimiterspace} 2}}  \\
   {0,} & {{{\bar \xi }_j} <  - {\Delta  \mathord{\left/
 {\vphantom {\Delta  2}} \right.
 \kern-\nulldelimiterspace} 2}}  \\
\end{array}} \right.}  \\
   {{N_ - } = N - {N_{\rm{ + }}}{\rm{ = }}2 - {N_{\rm{ + }}}}  \\
\end{array}} \right..
\end{equation}
The decision error probability of information number ${N_{\rm{ + }}}$ and ${N_ - }$ is given by
\begin{small}\begin{equation}\label{Eq_MF_pro8}
\begin{array}{l}
 {P_{MF}} =  \\
 \Pr \left( {{s_{1,j}} = 1,{s_{2,j}} =  - 1} \right)\Pr \left( {{N_{error}} = 1|{s_{1,j}} = 1,{s_{2,j}} =  - 1} \right) \\
  + \Pr \left( {{s_{1,j}} =  - 1,{s_{2,j}} = 1} \right)\Pr \left( {{N_{error}} = 1|{s_{1,j}} =  - 1,{s_{2,j}} = 1} \right) \\
 {\rm{ + }}\Pr \left( {{s_{1,j}} = 1,{s_{2,j}} = 1} \right)\Pr \left( {{N_{error}} = 1|{s_{1,j}} = 1,{s_{2,j}} = 1} \right) \\
 {\rm{ + }}\Pr \left( {{s_{1,j}} =  - 1,{s_{2,j}} =  - 1} \right)\Pr \left( {{N_{error}} = 1|{s_{1,j}} =  - 1,{s_{2,j}} =  - 1} \right) \\
 {\rm{ + }}\Pr \left( {{s_{1,j}} = 1,{s_{2,j}} = 1} \right)\Pr \left( {{N_{error}} = 2|{s_{1,j}} = 1,{s_{2,j}} = 1} \right) \\
 {\rm{ + }}\Pr \left( {{s_{1,j}} =  - 1,{s_{2,j}} =  - 1} \right)\Pr \left( {{N_{error}} = 2|{s_{1,j}} =  - 1,{s_{2,j}} =  - 1} \right) \\
 \end{array}.
\end{equation}\end{small}
The first term means that the transmitted information bits are ${s_{1,j}} = 1$ and ${s_{2,j}} =  - 1$ at $j$th symbol while the error number ${N_{error}} = {\rm{1}}$, i.e., the recovered bits could only be $\left( {{s_{1,j}} = 1,{s_{2,j}} = 1} \right)$ or $\left( {{s_{1,j}} =  - 1,{s_{2,j}} =  - 1} \right)$. The decision error probability of the first term is given as
\begin{equation}\label{Eq_MF_pro9}
\begin{array}{l}
 {P_{{M_1}}}{\rm{ = }}\Pr \left( {{N_{error}} = 1|{s_{1,j}} = 1,{s_{2,j}} =  - 1} \right) \\
 {\rm{ = }}\frac{1}{2}{\rm{erfc}}\left( {\frac{{{E_1} - {E_2} - \Xi {\rm{ + }}{\Delta  \mathord{\left/
 {\vphantom {\Delta  2}} \right.
 \kern-\nulldelimiterspace} 2}}}{{\sqrt 2 \sqrt {{E_1} + {E_2}} \sigma }}} \right), \\
 \end{array}
\end{equation}
where ${E_{\rm{1}}}$, ${E_{\rm{2}}}$ and $\Delta$ are given in Eq. (\ref{Eq_MF_pro1}) and Eq. (\ref{Eq_MF_pro6}), respectively, $\sigma $ denotes the standard deviation of the Gaussian noise, $\Xi $ is the sum of the ISI caused by the subcarrier itself, given by
\begin{align*}%\label{Eq_MF_pro10}
\Xi {\rm{ = }}\sum\limits_{n = 1}^{N = 2} {\sum\limits_{\scriptstyle m = 1 \hfill \atop
  \scriptstyle m \ne j \hfill}^M {{s_{n,m}}{I_{n,m}}} },
\end{align*}
and ${\rm{erfc}}\left(  \cdot  \right)$ is the complementary error function, defined as
\begin{align*}%\label{Eq_MF_pro11}
{\rm{erfc}}\left( x \right) = \frac{2}{{\sqrt \pi  }}\int\limits_x^\infty  {{e^{ - {t^2}}}} dt.
\end{align*}
The decision error probability of the second term is given as
\begin{equation}\label{Eq_MF_pro12}
\begin{array}{l}
 {P_{{M_2}}}{\rm{ = }}\Pr \left( {{N_{error}} = 1|{s_{1,j}} =  - 1,{s_{2,j}} = 1} \right) \\
 {\rm{ = }}\frac{1}{2}{\rm{erfc}}\left( {\frac{{ - {E_1} + {E_2} - \Xi {\rm{ + }}{\Delta  \mathord{\left/
 {\vphantom {\Delta  2}} \right.
 \kern-\nulldelimiterspace} 2}}}{{\sqrt 2 \sqrt {{E_1} + {E_2}} \sigma }}} \right). \\
 \end{array}
\end{equation}
The decision error probability of the third term is given as
\begin{equation}\label{Eq_MF_pro13}
\begin{array}{l}
 {P_{{M_3}}}{\rm{ = }}\Pr \left( {{N_{error}} = 1|{s_{1,j}} = 1,{s_{2,j}} = 1} \right) \\
 {\rm{       = }}\frac{1}{4}\left[ \begin{array}{l}
 {\rm{erfc}}\left( {\frac{{{E_1} + {E_2}{\rm{ + }}{B_1}{\rm{ + }}{B_2}{\rm{ + }}\Xi  - {\Delta  \mathord{\left/
 {\vphantom {\Delta  2}} \right.
 \kern-\nulldelimiterspace} 2}}}{{\sqrt 2 \sqrt {{E_1} + {E_2}} \sigma }}} \right) \\
  - {\rm{erfc}}\left( {\frac{{{E_1} + {E_2}{\rm{ + }}{B_1}{\rm{ + }}{B_2}{\rm{ + }}\Xi  + {\Delta  \mathord{\left/
 {\vphantom {\Delta  2}} \right.
 \kern-\nulldelimiterspace} 2}}}{{\sqrt 2 \sqrt {{E_1} + {E_2}} \sigma }}} \right) \\
 \end{array} \right]. \\
 \end{array}
\end{equation}
The decision error probability of the fourth term is given as
\begin{equation}\label{Eq_MF_pro14}
\begin{array}{l}
 {P_{{M_4}}}{\rm{ = }}\Pr \left( {{N_{error}} = 1|{s_{1,j}} =  - 1,{s_{2,j}} =  - 1} \right) \\
 {\rm{       = }}\frac{1}{4}\left[ \begin{array}{l}
 {\rm{erfc}}\left( {\frac{{ - {E_1} - {E_2} - {B_1} - {B_2}{\rm{ + }}\Xi  - {\Delta  \mathord{\left/
 {\vphantom {\Delta  2}} \right.
 \kern-\nulldelimiterspace} 2}}}{{\sqrt 2 \sqrt {{E_1} + {E_2}} \sigma }}} \right) \\
  - {\rm{erfc}}\left( {\frac{{ - {E_1} - {E_2} - {B_1} - {B_2}{\rm{ + }}\Xi  + {\Delta  \mathord{\left/
 {\vphantom {\Delta  2}} \right.
 \kern-\nulldelimiterspace} 2}}}{{\sqrt 2 \sqrt {{E_1} + {E_2}} \sigma }}} \right) \\
 \end{array} \right]. \\
 \end{array}
\end{equation}
The decision error probability of the fifth term is given as
\begin{equation}\label{Eq_MF_pro15}
\begin{array}{l}
 {P_{{M_5}}}{\rm{ = }}\Pr \left( {{N_{error}} = 2|{s_{1,j}} = 1,{s_{2,j}} = 1} \right) \\
 {\rm{ = }}\frac{1}{2}{\rm{erfc}}\left( {\frac{{{E_1}{\rm{ + }}{E_2}{\rm{ + }}{B_1}{\rm{ + }}{B_2}{\rm{ + }}\Xi {\rm{ + }}{\Delta  \mathord{\left/
 {\vphantom {\Delta  2}} \right.
 \kern-\nulldelimiterspace} 2}}}{{\sqrt 2 \sqrt {{E_1} + {E_2}} \sigma }}} \right). \\
 \end{array}
\end{equation}
The decision error probability of the sixth term is given as
\begin{equation}\label{Eq_MF_pro16}
\begin{array}{l}
 {P_{{M_6}}}{\rm{ = }}\Pr \left( {{N_{error}} = 2|{s_{1,j}} =  - 1,{s_{2,j}} =  - 1} \right) \\
 {\rm{ = }}\frac{1}{2}{\rm{erfc}}\left( {\frac{{ - {E_1} - {E_2} - {B_1} - {B_2}{\rm{ + }}\Xi {\rm{ + }}{\Delta  \mathord{\left/
 {\vphantom {\Delta  2}} \right.
 \kern-\nulldelimiterspace} 2}}}{{\sqrt 2 \sqrt {{E_1} + {E_2}} \sigma }}} \right). \\
 \end{array}
\end{equation}
Substituting Eqs. (\ref{Eq_MF_pro9}-\ref{Eq_MF_pro16}) into Eq. (\ref{Eq_MF_pro8}) we get the decision error probability of number ${N_{\rm{ + }}}$ and ${N_ - }$ as
\begin{equation}\label{Eq_MF_pro17}
{P_{CPOMF}} = \frac{1}{4}\left( {{P_{{M_1}}}{\rm{ + }}{P_{{M_2}}}{\rm{ + }}{P_{{M_3}}}{\rm{ + }}{P_{{M_4}}}{\rm{ + }}{P_{{M_5}}}{\rm{ + }}{P_{{M_6}}}} \right).
\end{equation}
The CPOCF bank outputs are
\begin{equation}\label{Eq_CPOF_pro1}
\begin{array}{l}
 {y_1}\left( t \right) = \int\limits_{ - \infty }^\infty  {R\left( \tau  \right){o_1}\left( {t - \tau } \right)} d\tau  \\
 {\rm{ = }}\sum\limits_{m = 1}^M {{s_{1,m}}} \int\limits_{ - \infty }^\infty  {{p_1}\left( \tau  \right){o_1}\left( {\tau  - t + m/f} \right)} d\tau  \\
  + \sum\limits_{m = 1}^M {{s_{2,m}}} \int\limits_{ - \infty }^\infty  {{p_2}\left( \tau  \right){o_2}\left( {\tau  - t + m/f} \right)} d\tau  \\
  + \int\limits_{ - \infty }^\infty  {w\left( \tau  \right){o_1}\left( {\tau  - t} \right)} d\tau  \\
 \end{array}
\end{equation}
and
\begin{equation}\label{Eq_CPOF_pro2}
\begin{array}{l}
 {y_2}\left( t \right) = \int\limits_{ - \infty }^\infty  {R\left( \tau  \right){o_2}\left( {t - \tau } \right)} d\tau  \\
 {\rm{ = }}\sum\limits_{m = 1}^M {{s_{2,m}}} \int\limits_{ - \infty }^\infty  {{p_2}\left( \tau  \right){o_2}\left( {\tau  - t + m/f} \right)} d\tau  \\
  + \sum\limits_{m = 1}^M {{s_{1,m}}} \int\limits_{ - \infty }^\infty  {{p_1}\left( \tau  \right){o_2}\left( {\tau  - t + m/f} \right)} d\tau  \\
  + \int\limits_{ - \infty }^\infty  {w\left( \tau  \right){o_2}\left( {\tau  - t} \right)} d\tau.  \\
 \end{array}
\end{equation}
By the same way, the $j$th $\left( {j = {\rm{1}}, \ldots ,M} \right)$ sampling points ${y_{1,j}}$ and ${y_{2,j}}$ from ${y_1}\left( t \right)$ and ${y_2}\left( t \right)$ can be expressed as
\begin{equation}\label{Eq_CPOF_pro3}
\begin{array}{l}
 {y_{1,j}}{\rm{ = }}{s_{1,j}}{Q_1} + \sum\limits_{m = 1}^{j - 1} {{s_{1,m}}{A_{1,m}}}  + \sum\limits_{m = j + 1}^M {{s_{1,m}}{B_{1,m}}}  \\
  + {s_{2,m}}{C_1} + \sum\limits_{m = 1}^{j - 1} {{s_{2,m}}{X_{1,m}}}  + \sum\limits_{m = j + 1}^M {{s_{2,m}}{Y_{1,m}}}  + {W_{{C_1}}} \\
 \end{array}
\end{equation}
and
\begin{equation}\label{Eq_CPOF_pro4}
\begin{array}{l}
 {y_{2,j}}{\rm{ = }}{s_{2,j}}{Q_2} + \sum\limits_{m = 1}^{j - 1} {{s_{2,m}}{A_{2,m}}}  + \sum\limits_{m = j + 1}^M {{s_{2,m}}{B_{2,m}}}  \\
  + {s_{1,m}}{C_2} + \sum\limits_{m = 1}^{j - 1} {{s_{1,m}}{X_{2,m}}}  + \sum\limits_{m = j + 1}^M {{s_{1,m}}{Y_{2,m}}}  + {W_{{C_2}}}, \\
 \end{array}
\end{equation}
where ${Q_n}\left( {n = {\rm{1}},{\rm{2}}} \right)$ is the expected energy and equal to the correlation between the basis function and the corresponding CPOCF, given by
\begin{equation}\label{Eq_CPOF_pro5}
{Q_n} = \begin{array}{*{20}{c}}
   {\frac{{\left( {1 - {e^{ - \beta }}} \right)}}{2}\left[ {{\Delta _{n,1}} - \frac{{2\beta }}{{{\beta ^2} + \omega _n^2}}} \right],} & {n = 1,2}  \\
\end{array},
\end{equation}
${A_{n,m}}$ and ${B_{n,m}}$ are the ISI caused by past and future symbols of the subcarrier itself, given as
\begin{equation}\label{Eq_CPOF_pro6}
\begin{array}{*{20}{c}}
   \begin{array}{l}
 {A_{n,m}} =  \\
 \left[ {{\Delta _{n,1}} + \frac{{2\beta }}{{{\beta ^2} + \omega _n^2}}} \right]\frac{{\left( {2 - {e^{ - \beta }} - {e^\beta }} \right){e^{ - \beta m}}}}{4} \\
 {\rm{ + }}{\left( {1 - {e^{ - \beta }}} \right)^2}{e^{\beta \left( {1 - m} \right)}}\frac{{2\beta }}{{{\beta ^2} + \omega _n^2}}, \\
 \end{array} & {m = 1,2,...,j - 1}  \\
\end{array}
\end{equation}
and
\begin{equation}\label{Eq_CPOF_pro7}
\begin{array}{*{20}{c}}
   \begin{array}{l}
 {B_{n,m}} =  \\
 \frac{1}{4}\left( {{\Delta _{n,1}} + \frac{{2\beta }}{{{\beta ^2} + \omega _n^2}}} \right) \\
 \left( {1 - {e^{ - \beta }}} \right)\left( {{e^{ - \beta m}} - {e^{\beta \left( {1 - m} \right)}}} \right), \\
 \end{array} & {m = j + 1,j + 2,...,M}  \\
\end{array}.
\end{equation}
${C_n}$ is the ISI caused by the other subcarriers at the $j$th sampling instant and is equal to the cross-correlation between the basis function and the other CPOCF, given by
\begin{equation}\label{Eq_CPOF_pro8}
{C_n} = \left( {1 - {e^{ - \beta }}} \right)\left[ {{\Delta _{n,2}} - \frac{{2\beta }}{{{\beta ^2} + \omega _n^2}}} \right].
\end{equation}
${X_{n,m}}$ and ${Y_{n,m}}$ are the ISI caused by past and future symbols of the other subcarriers, given as
\begin{equation}\label{Eq_CPOF_pro9}
\begin{array}{*{20}{c}}
   \begin{array}{l}
 {X_{n,m}} = {\left( {1 - {e^{ - \beta }}} \right)^2}{e^{\beta \left( {1 - m} \right)}}\frac{{2\beta }}{{{\beta ^2} + \omega _n^2}} \\
 {\rm{ + }}{\Delta _{n,2}}\frac{{\left( {2 - {e^{ - \beta }} - {e^\beta }} \right){e^{ - \beta m}}}}{2} \\
 \end{array} & {m = 1,2,...,j - 1}  \\
\end{array}
\end{equation}
and
\begin{equation}\label{Eq_CPOF_pro10}
\begin{array}{l}
 {Y_{n,m}} = {\Delta _{n,2}}\left( \begin{array}{l}
 \frac{{{e^{ - \beta m}}{{\left( {1 - {e^{ - \beta }}} \right)}^2}}}{2} -  \\
 \frac{{\left( {1 - {e^{ - \beta }}} \right)\left( {{e^{\beta \left( {1 - m} \right)}} - {e^{ - \beta \left( {1 - m} \right)}}} \right)}}{2} \\
 \end{array} \right), \\
 {\rm{   \qquad\qquad\qquad\qquad\qquad                                        }}m = j + 1,j + 2,...,M. \\
 \end{array}
\end{equation}
The bit error probability of CPOCF is given by
\begin{small}\begin{equation}\label{Eq_CPOF_pro11}
\begin{array}{l}
 {P_{CPOCF}} =  \\
 \Pr \left( {{s_{1,j}} = 1,{s_{2,j}} =  - 1} \right)\Pr \left( {{y_{1,j}} < {y_{2,j}}|{s_{1,j}} = 1,{s_{2,j}} =  - 1} \right) \\
  + \Pr \left( {{s_{1,j}} =  - 1,{s_{2,j}} = 1} \right)\Pr \left( {{y_{1,j}} \ge {y_{2,j}}|{s_{1,j}} =  - 1,{s_{2,j}} = 1} \right) \\
 \end{array}.
\end{equation}\end{small}
It should be noticed that $\Pr \left( {{s_{1,j}} = 1,{s_{2,j}} = 1} \right)$ and $\Pr \left( {{s_{1,j}} =  - 1,{s_{2,j}} =  - 1} \right)$ are not considered in Eq. (\ref{Eq_CPOF_pro11}). This is because the information bits can be recovered without CPOCF if ${N_{\rm{ + }}}{\rm{ = }}N{\rm{ = }}2$ (or ${N_ - }{\rm{ = }}N{\rm{ = }}2$) in Eq. (\ref{Eq_MF_pro7}), which means that the recovered bits are $\left( {{s_{1,j}},{s_{2,j}}} \right) = \left( { + 1, + 1} \right)$ (or $\left( {{s_{1,j}},{s_{2,j}}} \right) = \left( { - 1, - 1} \right)$) and the CPOCF does not work in this case.

The bit error probability of the first term is given as
\begin{equation}\label{Eq_CPOF_pro12}
\begin{array}{l}
 {P_{{C_1}}} = \Pr \left( {{y_{1,j}} < {y_{2,j}}|{s_{1,j}} = 1,{s_{2,j}} =  - 1} \right) \\
 {\rm{ = }}\frac{1}{2}{\rm{erfc}}\left( {\frac{{{Q_1} - {Q_2} - {C_1} - {C_2} + {\Phi _1} - {\Phi _2}}}{{\sqrt 2 \sqrt {{Q_1} + {Q_2}} \sigma }}} \right), \\
 \end{array}
\end{equation}
where ${Q_n}$ and ${C_n}$ are given in Eq. (\ref{Eq_CPOF_pro5}) and Eq. (\ref{Eq_CPOF_pro8}), ${\Phi _n}$ denotes the sum of the ISI caused by the past and future symbol in both subcarriers, given as
\begin{align*}%\label{Eq_CPOF_pro13}
\begin{array}{*{20}{c}}
   \begin{array}{l}
 {\Phi _n}{\rm{ = }}\sum\limits_{m = 1}^{j - 1} {{s_{n,m}}{A_{n,m}}}  + \sum\limits_{m = j + 1}^M {{s_{n,m}}{B_{n,m}}}  +  \\
 \sum\limits_{m = 1}^{j - 1} {{s_{3 - n,m}}{X_{n,m}}}  + \sum\limits_{m = j + 1}^M {{s_{3 - n,m}}{Y_{n,m}}} , \\
 \end{array} & {n = 1,2}  \\
\end{array}.
\end{align*}
The bit error probability of the second term is given as
\begin{equation}\label{Eq_CPOF_pro14}
\begin{array}{l}
 {P_{{C_2}}} = \Pr \left( {{y_{1,j}} \ge {y_{2,j}}|{s_{1,j}} =  - 1,{s_{2,j}} = 1} \right) \\
 {\rm{ = }}\frac{1}{2}{\rm{erfc}}\left( {\frac{{ - {Q_1} + {Q_2} - {C_1} - {C_2} - {\Phi _1} + {\Phi _2}}}{{\sqrt 2 \sqrt {{Q_1} + {Q_2}} \sigma }}} \right) \\
 \end{array}.
\end{equation}
By substituting Eqs. (\ref{Eq_CPOF_pro12}-\ref{Eq_CPOF_pro14}) into Eq. (\ref{Eq_CPOF_pro11}) we get the bit error probability of CPOCF as
\begin{equation}\label{Eq_CPOF_pro15}
{P_{CPOCF}} = \frac{1}{4}{P_{{C_1}}} + \frac{1}{4}{P_{{C_2}}}.
\end{equation}
The bit error probability of the proposed CPOCMA is given as
\begin{equation}\label{Eq_CPOF_pro16}
\begin{array}{l}
 {P_e} = {P_{CPOMF}}{P_{CPOCF}} \\
  + {{\bar P}_{CPOMF}}{P_{CPOCF}} + {P_{CPOMF}}{{\bar P}_{CPOCF}} \\
  = {P_{CPOMF}}{P_{CPOCF}} + \left( {1 - {P_{CPOMF}}} \right){P_{CPOCF}} \\
  + {P_{CPOMF}}\left( {1 - {P_{CPOCF}}} \right), \\
 \end{array}
\end{equation}
where ${P_{CPOMF}}$ and ${P_{CPOCF}}$ are given in Eq. (\ref{Eq_MF_pro17}) and Eq. (\ref{Eq_CPOF_pro15}), respectively. To this end, the BER of 2 sub-carriers of the proposed CPOCMA system is derived as Eq. (\ref{Eq_CPOF_pro16}).

\section{Simulation Results}
\subsection{The single-user scenario}
\begin{figure}[!t]
\centering
\includegraphics[width=3.5in]{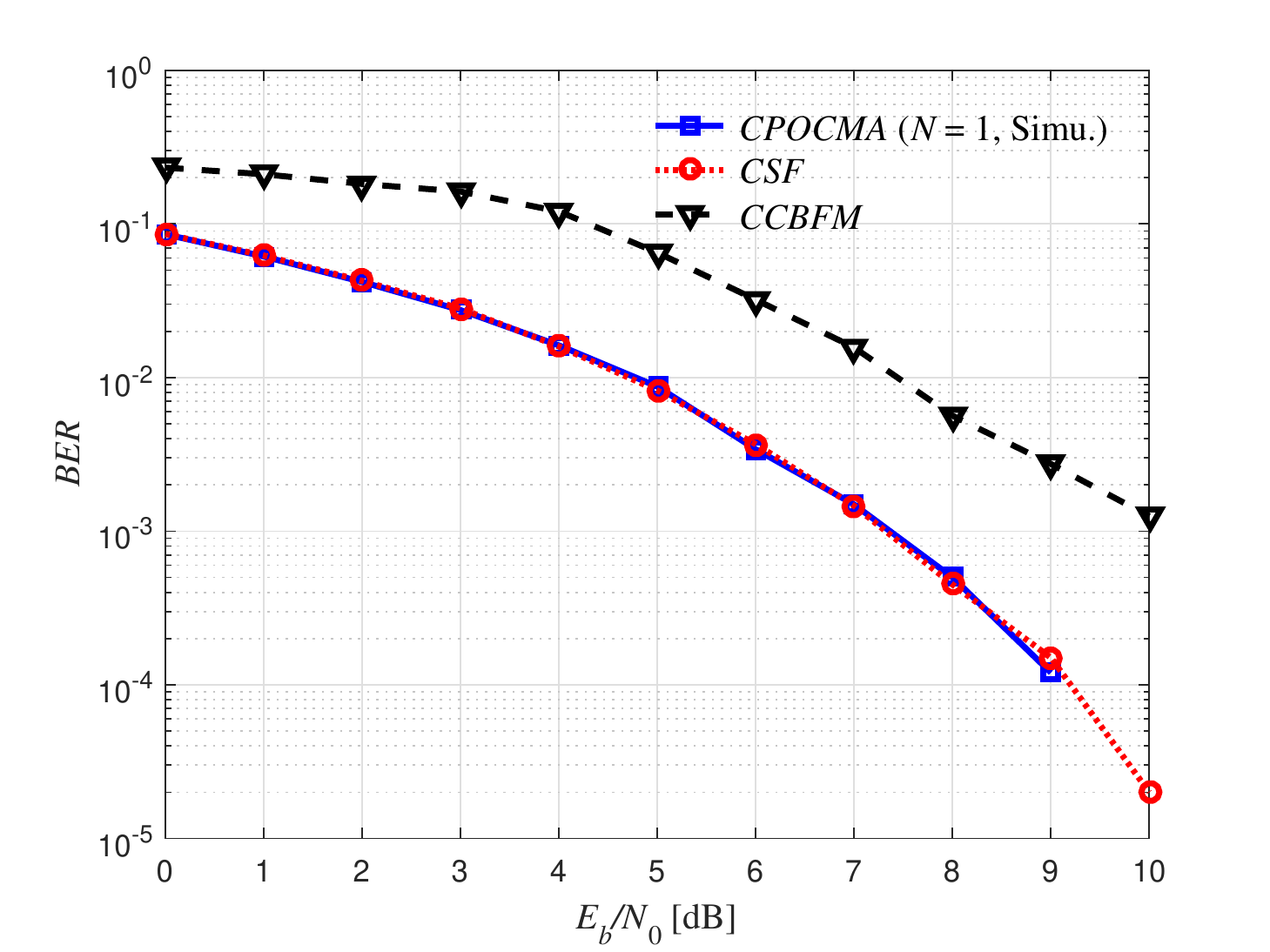}
\caption{Simulation performance comparison for single subcarrier under an AWGN channel.}
\label{Fig_AWGN}
\end{figure}

In this subsection, the BER performance comparison of the proposed CPOCMA system and two novel chaotic symbolic dynamics modulations, i.e., the CSF \cite{Bai2020_Radio} and CCBFM \cite{Pappu2020_Simultaneous} systems are analyzed for single-user case. The simulated BER curves over an AWGN channel are given in Fig. \ref{Fig_AWGN}. The initial frequency $f$ = 2.5MHz, the base frequency of the CSF and CPOCMA is ${f_1}$ = 2.5MHz, the up-carrier frequency is ${f_c}$ = 5MHz and the sampling frequency ${f_s}$ = 40 MHz in the CSF, CPOCMA and CCBFM schemes. The blue solid line with square marks, the red dotted line with circle marks and the black dashed line with triangle marks are the simulation BER curves of the CPOCMA, the CSF and the CCBFM, respectively. The CCBFM shows the worst BER performance since it is lack of ability to resist noise interference without a matched filter as done by CSF and CPOCMA. The CPOCMA shows the same BER performance as the CSF, because its modulation scheme is equivalent to CSF in single-user scenario. Moreover, the CSF and CCBFM can only transmit single data stream, while CPOCMA can transmit multiple sequences and achieve higher BTR, meanwhile, it can be further extended to multi-subcarrier multi-user transmission.
\subsection{The multi-user scenario}
\begin{figure*}
\begin{minipage}[!t]{0.5\linewidth}
\centering
  \subfigure[ ]{
    \label{Fig_MULTI_AWGN} %% label for first subfigure
    \includegraphics[scale=0.54,trim= 0 0 0 0]{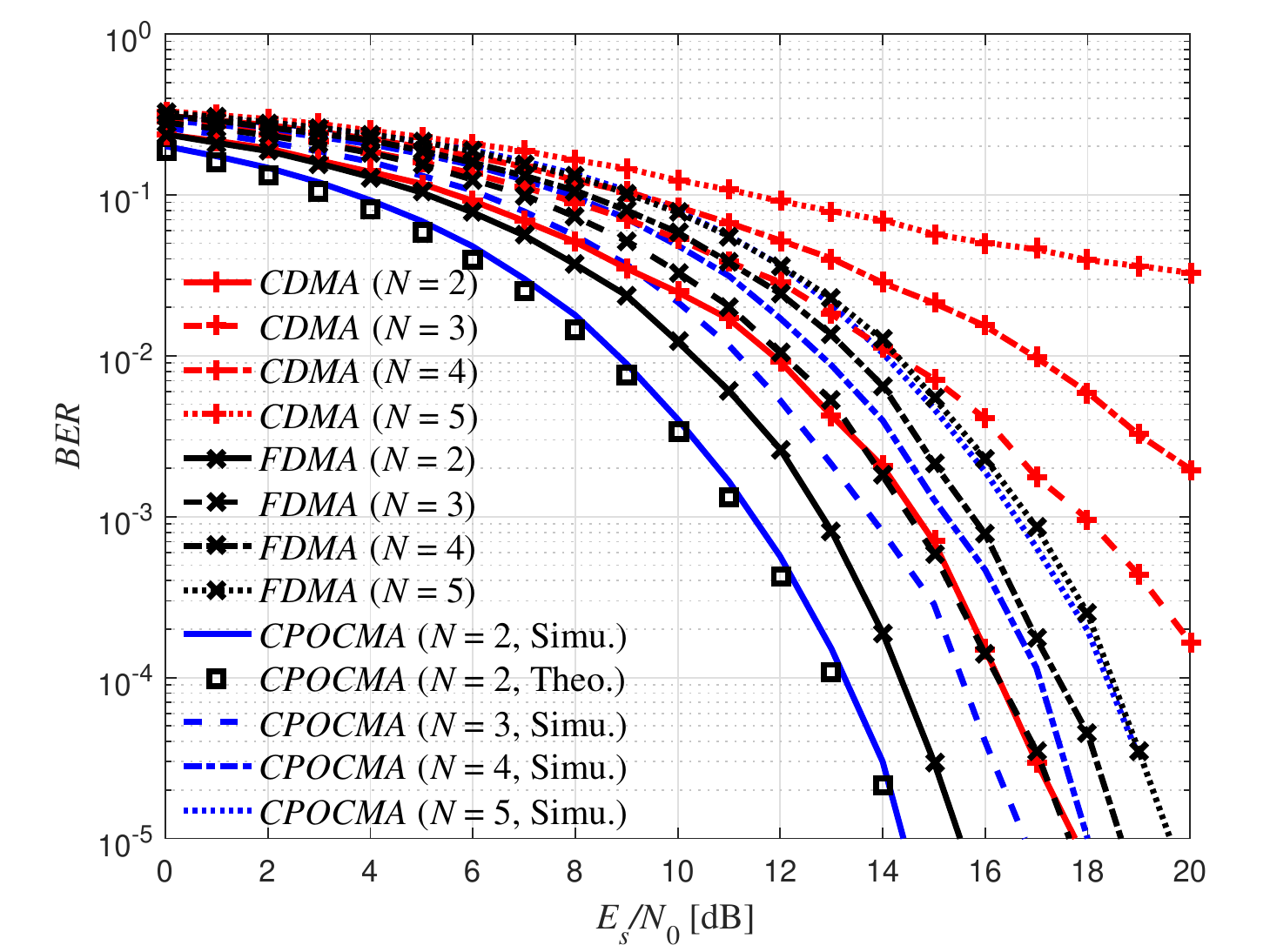}}
  \end{minipage}%
  \begin{minipage}[!t]{0.5\linewidth}
 \centering
  \subfigure[ ]{
    \label{Fig_MULTI_multi} %% label for second subfigure
    \includegraphics[scale=0.57,trim= 0 0 0 0]{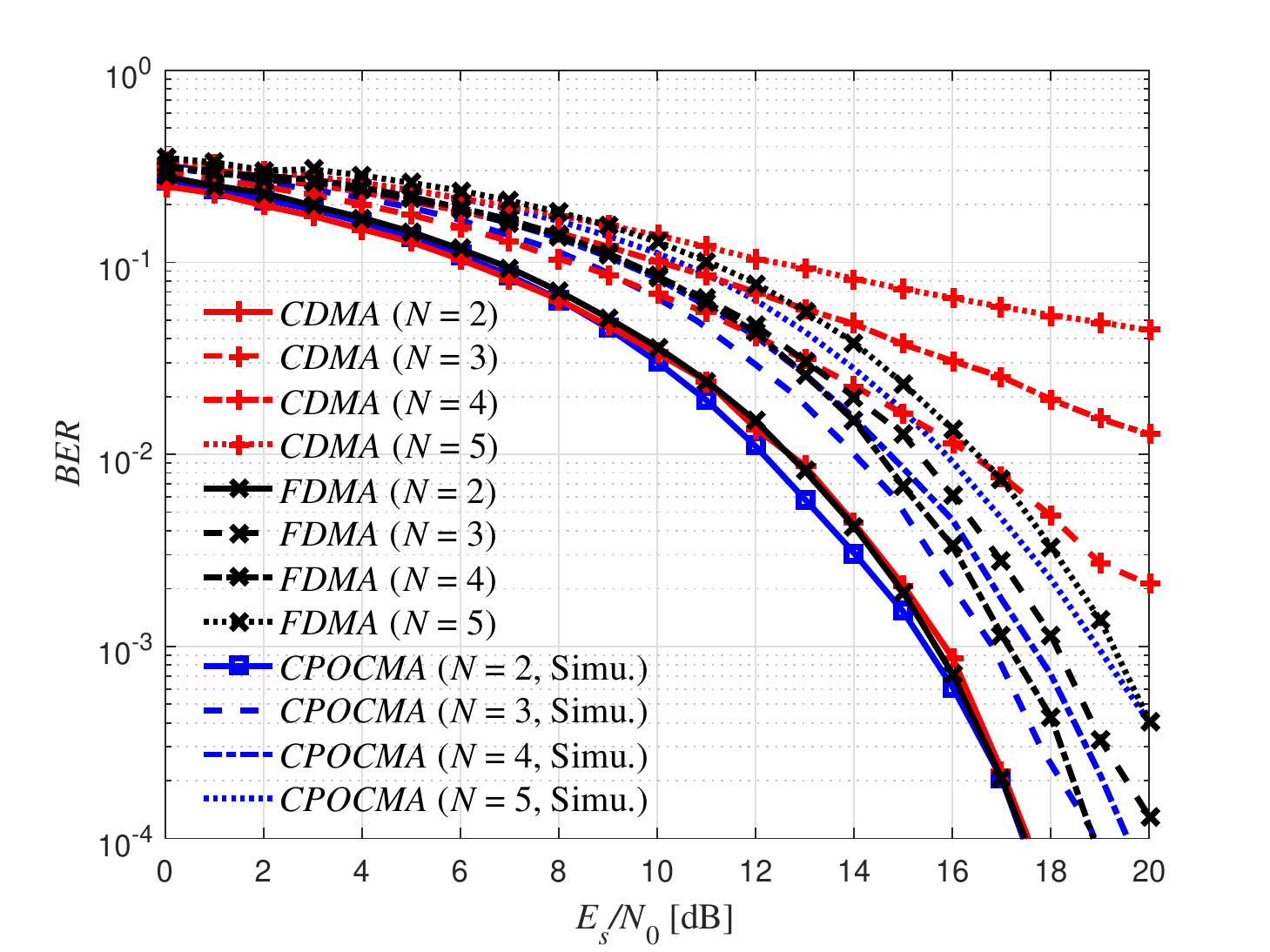}}
    \end{minipage}%
    \\
    \caption{Simulation BER comparisons for multi-subcarriers. (a) The simulation under an AWGN channel; (b) the simulation under a three-ray wireless channel with average power gains [0.7, 0.3, 0.1]dB and excess delay [0, 0.1, 0.125]$\mu s$.}
  \label{Fig_multi_subcarriers} %% label for entire figure
\end{figure*}

The performance comparison of the CPOCMA, CDMA and FDMA for multi-user case are analyzed in this part, where the simulations are performed with different number of subcarriers under an AWGN channel and under a wireless channel, respectively. The BER curves are given in Fig. \ref{Fig_multi_subcarriers}. In the CPOCMA, the initial frequency $f$ = 0.3125MHz and the base frequencies are set from $f$ to $Nf$, where $N$ = 2, 3, 4, 5. In the CDMA and FDMA, the oversampling rate in one symbol period is 128 and the data transmission rate of CDMA, FDMA and CPOCMA are the same, and the sampling frequency ${f_s}$ = 40MHz. Figure \ref{Fig_MULTI_multi} shows the simulation results in white noise channel, from Fig. \ref{Fig_MULTI_AWGN}, we know that there is an excellent match between simulated result and theoretical BER expression of the proposed CPOCMA for $N$ = 2 subcarriers. It shows Multiple Access Interference (MAI) caused by the additional users introduces a clear performance degradation in the multiuser case, and the performance degradation of CDMA is more serious than that of FDMA and the proposed CPOCMA. With the increasing number of users, the proposed CPOCMA shows similar BER performance as that of FDMA, but occupies less frequency spectrum, which will be given latterly in Sec. IV-C. Therefore, the proposed CPOCMA has better spectrum efficiency as compared to FDMA, while the proposed CPOCMA has less MAI as compared to CDMA under the same data transmission rate.

The performance of CPOCMA, CDMA and FDMA are presented and evaluated in Fig. \ref{Fig_MULTI_multi} for the same system parameters as these in Fig. \ref{Fig_MULTI_AWGN} under a three-ray wireless channel model, where the channel statistical parameters are average power gains [0.7, 0.2, 0.1]dB with excess delay [0, 0.1, 0.125]$\mu s$, respectively. The results obtained in Fig. \ref{Fig_MULTI_multi} confirm the fact that the system performance degrades rapidly with the increase of MAI as seen from Fig. \ref{Fig_MULTI_AWGN}. The CPOCMA, CDMA and FDMA show the similar BER performance when $N$ = 2, however, the performance gap increases as the number of users increase.

\subsection{Bandwidth efficiency and throughput analysis}
\begin{figure*}
\begin{minipage}[!t]{0.5\linewidth}
\centering
  \subfigure[ ]{
    \label{Fig_bandwidth} %% label for first subfigure
    \includegraphics[scale=0.55,trim= 0 0 0 0]{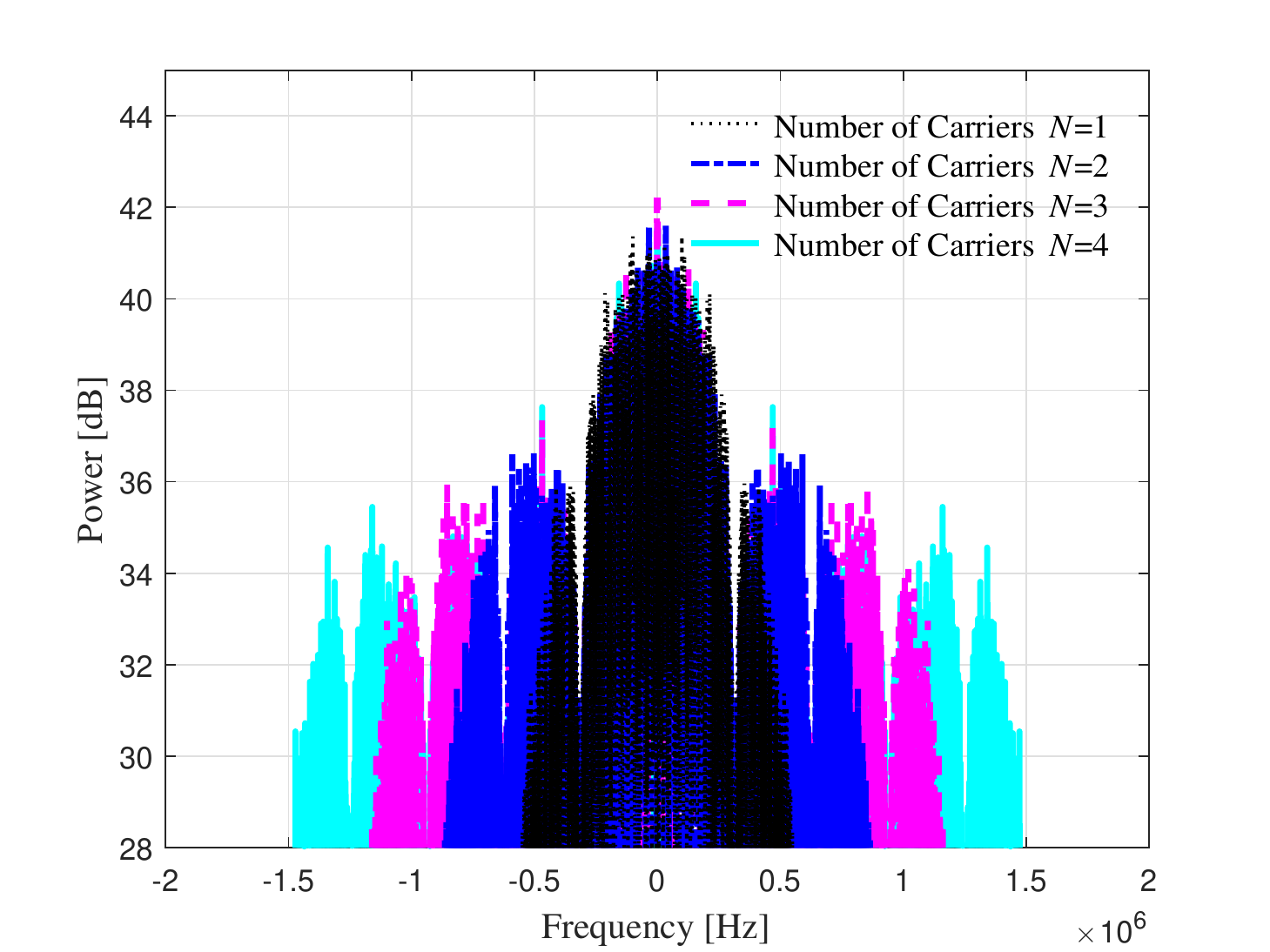}}
  \end{minipage}%
  \begin{minipage}[!t]{0.5\linewidth}
 \centering
  \subfigure[ ]{
    \label{Fig_throughout} %% label for second subfigure
    \includegraphics[scale=0.55,trim= 0 0 0 0]{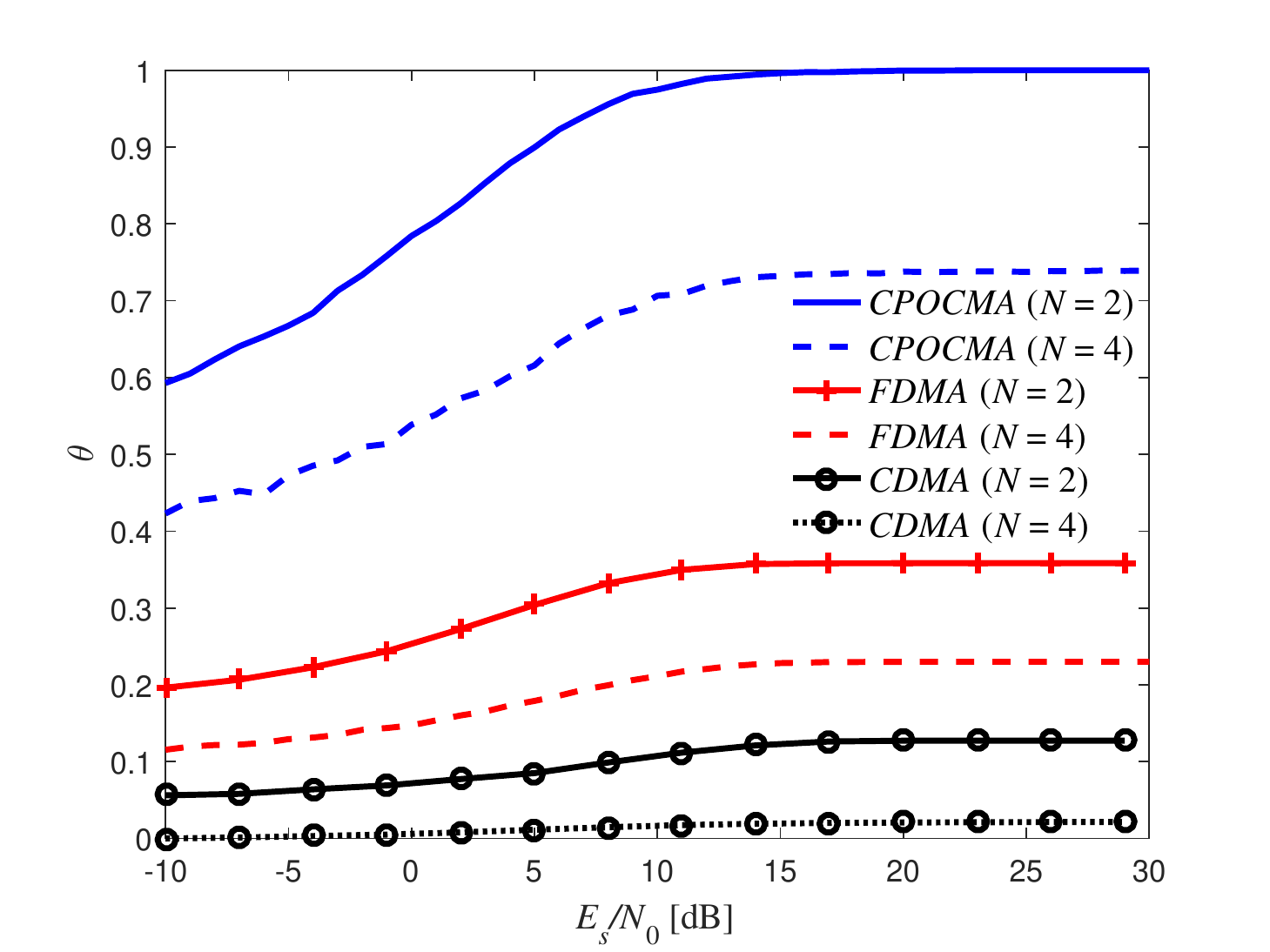}}
    \end{minipage}%
    \\
    \caption{The bandwidth efficiency analysis of the CPOCMA. (a) The bandwidth of the transmitted signal with $N$ = 1, 2, 3, 4 subcarriers for CPOCMA; (b) the throughput per frequency comparison between the CPOCMA, FDMA and CDMA.}
  \label{Fig_efficiency} %% label for entire figure
\end{figure*}
In order to verify that the proposed CPOCMA scheme has higher spectrum efficiency, the frequency spectrums of different carrier number are shown in Fig. \ref{Fig_bandwidth} with initial frequency $f$ = 0.3125MHz and base frequencies ${f_1}$ to ${f_4}$ are set from $f$ to ${4f}$. It is noticed that the spectrum increases with the increasing number of the subcarriers, and the increment is about half of the bandwidth for single carrier. The signal bandwidth when $N$ = 1 is approximately $B \approx 4f$, then the signal bandwidth of CPOCMA with $N$ subcarriers is approximately
\begin{align*}%\label{Eq_CPOF_pro13}
B_N^{CPOCMA} \approx B + \frac{B}{2}\left( {N - 1} \right) = 2fN + 2f,
\end{align*}
while the signal bandwidth of FDMA and CDMA are $B_N^{FDMA}{\rm{ = 2}}N\tilde B$ and $B_N^{CDMA} = P\tilde B$, respectively, where $P$ is the spreading gain of the CDMA and $\tilde B$ is the bandwidth of signal generated by conventional shaping forming filter, such as Square Root Raised Cosine (SRRC) filter. Assume that the bandwidth $\tilde B{\rm{ = }}B$, then $B_N^{FDMA}{\rm{ = }}8Nf \gg B_N^{CPOCMA}$, especially with the increase of subcarrier number $N$. The bandwidth of CDMA is proportional to the spreading gain, i.e., $B_N^{CDMA} = 4Pf \gg B_N^{CPOCMA}$ with the increase of spreading gain. The bandwidth efficiency of CPOCMA shows significantly superiority as compared to FDMA and CDMA.

A throughput per frequency is defined as the $\theta  = {\Theta  \mathord{\left/
 {\vphantom {\Theta  {{B_N}}}} \right.
 \kern-\nulldelimiterspace} {{B_N}}}$, $\left( {{B_N} \in \left\{ {B_N^{FDMA},B_N^{CPOCMA},B_N^{CDMA}} \right\}} \right)$, where $\Theta $ is the throughput defined as the ratio between the successfully received bits and the transmitted bits \cite{Ma2020_Design}. The normalized throughputs per frequency of CPOCMA, CDMA and FDMA under the wireless radio channel are given in Fig. \ref{Fig_throughout}, where the channel parameters are the same as that in Fig. \ref{Fig_MULTI_multi}. It shows the superiority of the CPOCMA in throughput per frequency compared with FDMA and CDMA, where $P$ = 16 in CDMA.

\section{The Experimental Configuration and Evaluation}
\begin{figure*}
\begin{minipage}[!t]{0.5\linewidth}
\centering
  \subfigure[ ]{
    \label{Fig_setup_photo} %% label for first subfigure
    \includegraphics[scale=0.4,trim= 0 0 0 0]{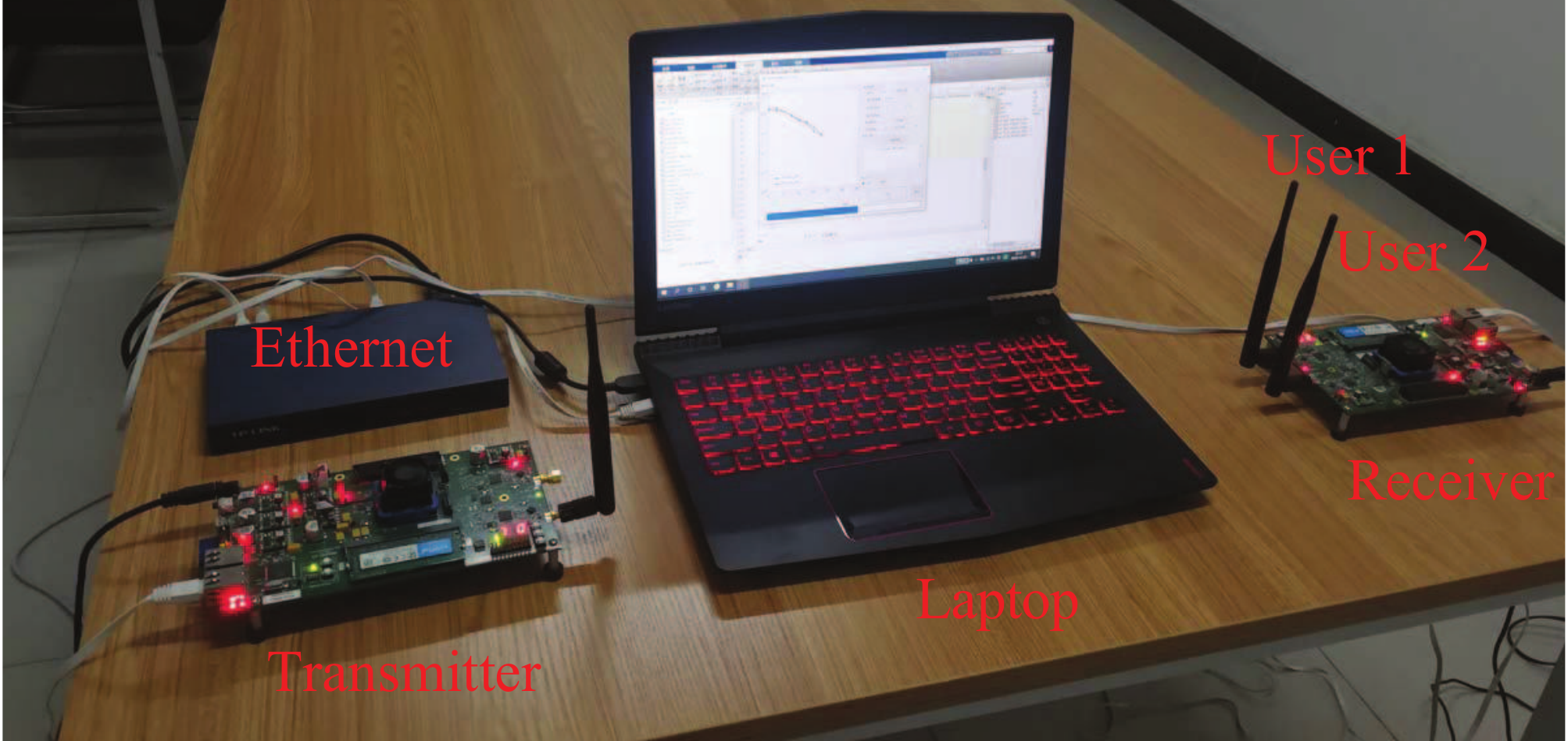}}
  \end{minipage}%
  \begin{minipage}[!t]{0.5\linewidth}
 \centering
  \subfigure[ ]{
    \label{Fig_frame_str} %% label for second subfigure
    \includegraphics[scale=0.45,trim= 0 0 0 0]{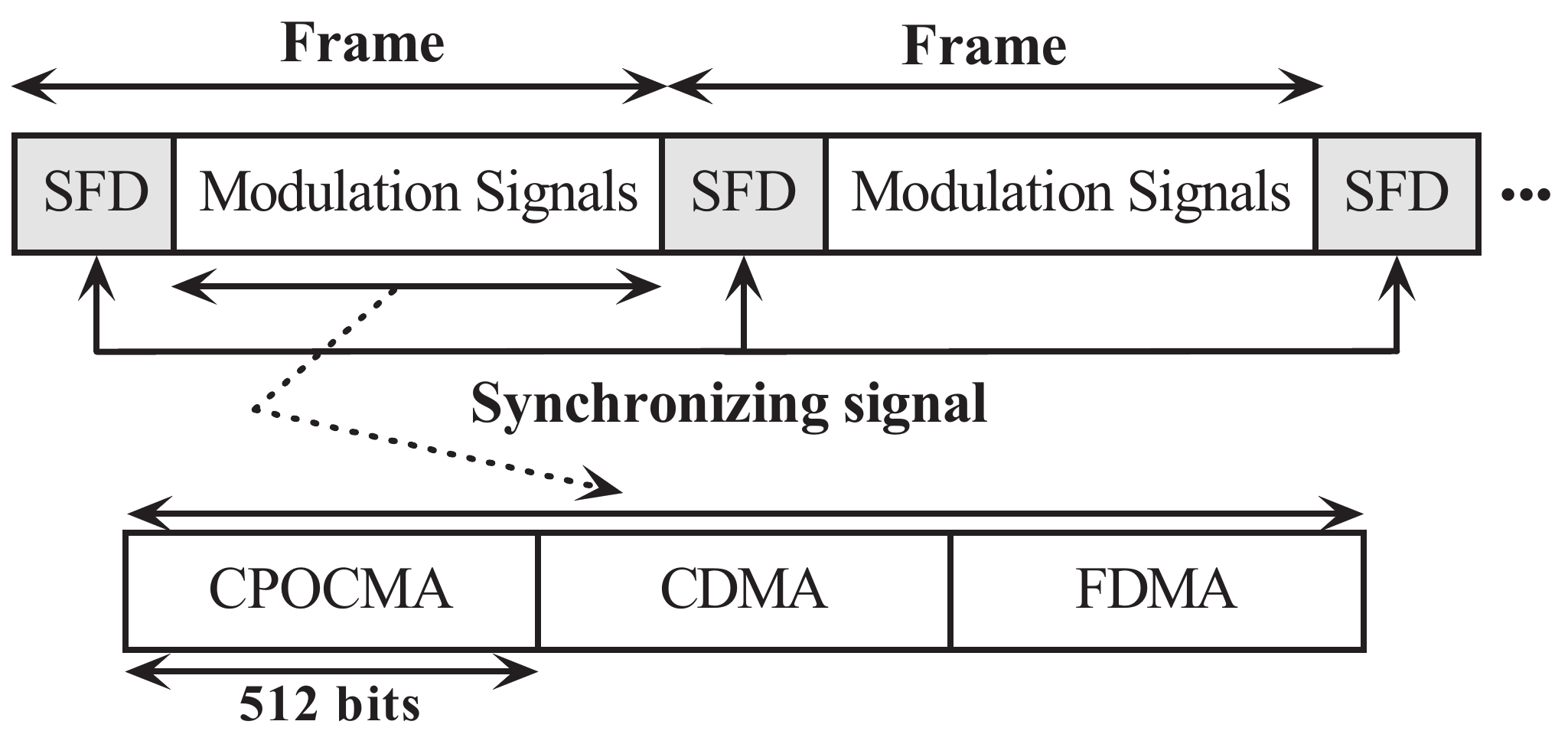}}
    \end{minipage}%
    \\
    \caption{The communication system configuration for single path channel. (a) The experiment hardware setup photo; (b) The frame structure of the transmitted signal containing SFD, modulated signals for CPOCMA, CDMA and FDMA.}
  \label{Fig_conf} %% label for entire figure
\end{figure*}

\begin{figure}
\begin{minipage}[!t]{0.5\linewidth}
\centering
  \subfigure[ ]{
    \label{Fig_exper_orig} %% label for first subfigure
    \includegraphics[scale=0.35,trim= 0 0 0 0]{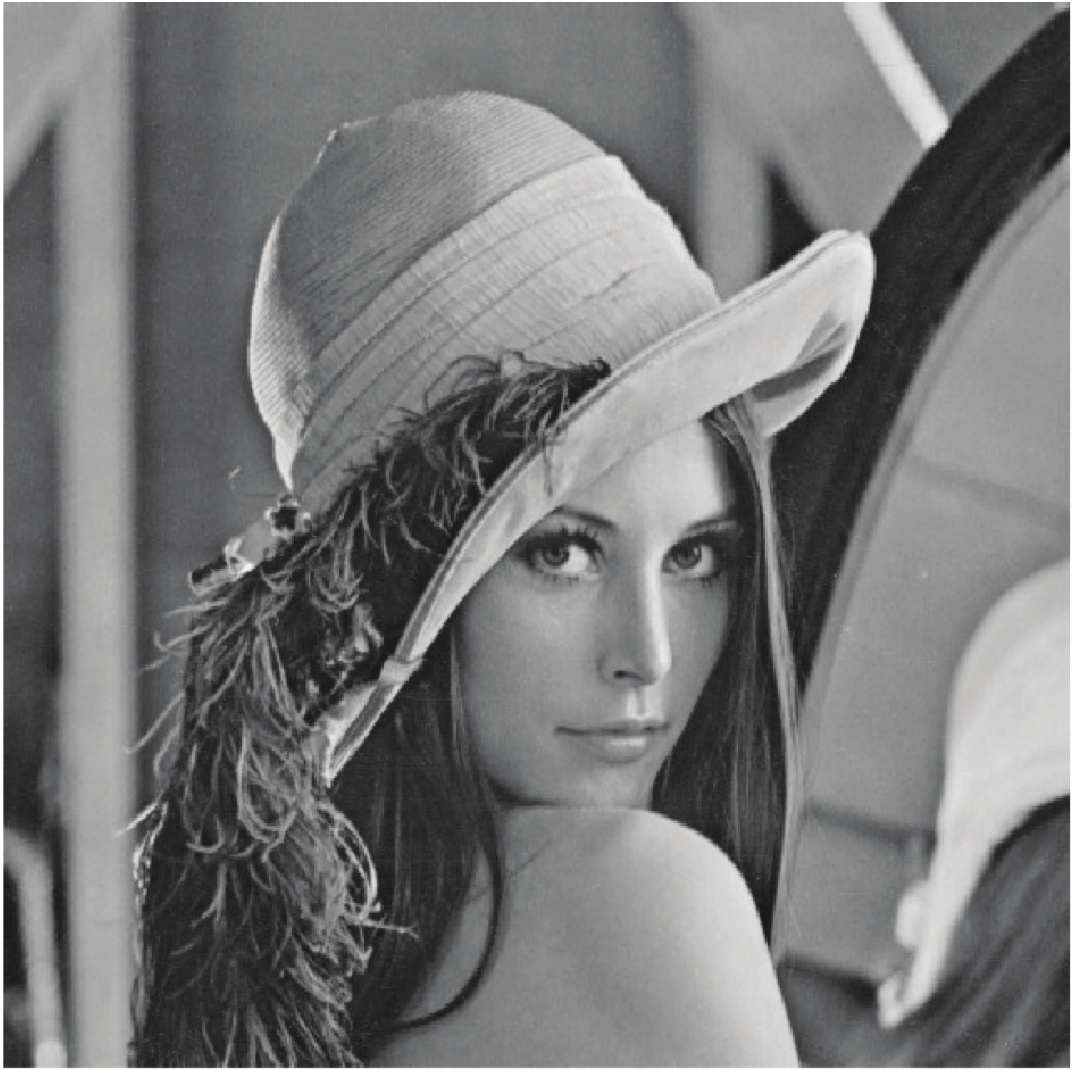}}
  \end{minipage}%
  \begin{minipage}[!t]{0.5\linewidth}
 \centering
  \subfigure[ ]{
    \label{Fig_exper_CPOCMA} %% label for second subfigure
    \includegraphics[scale=0.35,trim= 0 0 0 0]{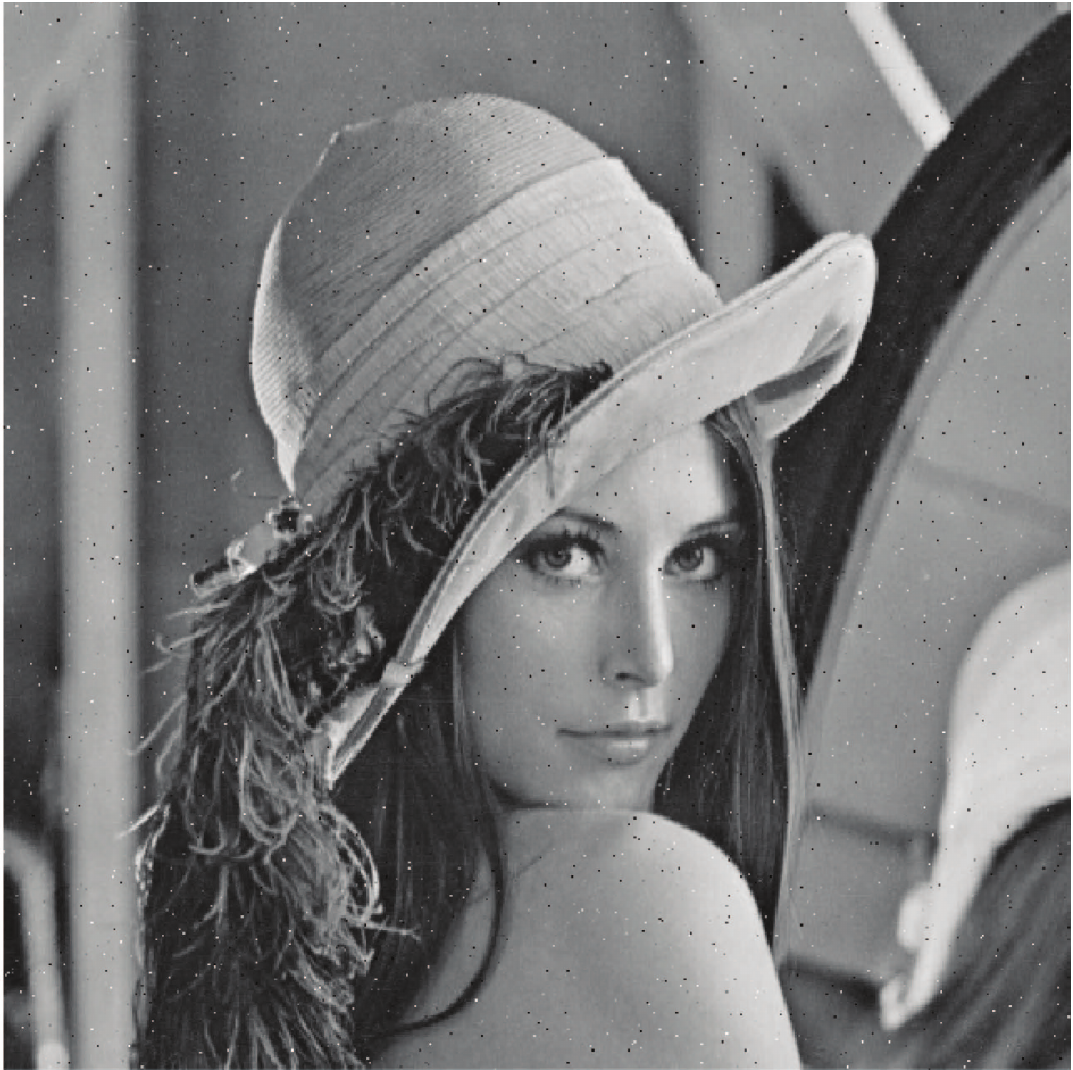}}
    \end{minipage}%
    \\
\begin{minipage}[!t]{0.5\linewidth}
\centering
  \subfigure[ ]{
    \label{Fig_exper_FDMA} %% label for first subfigure
    \includegraphics[scale=0.35,trim= 0 0 0 0]{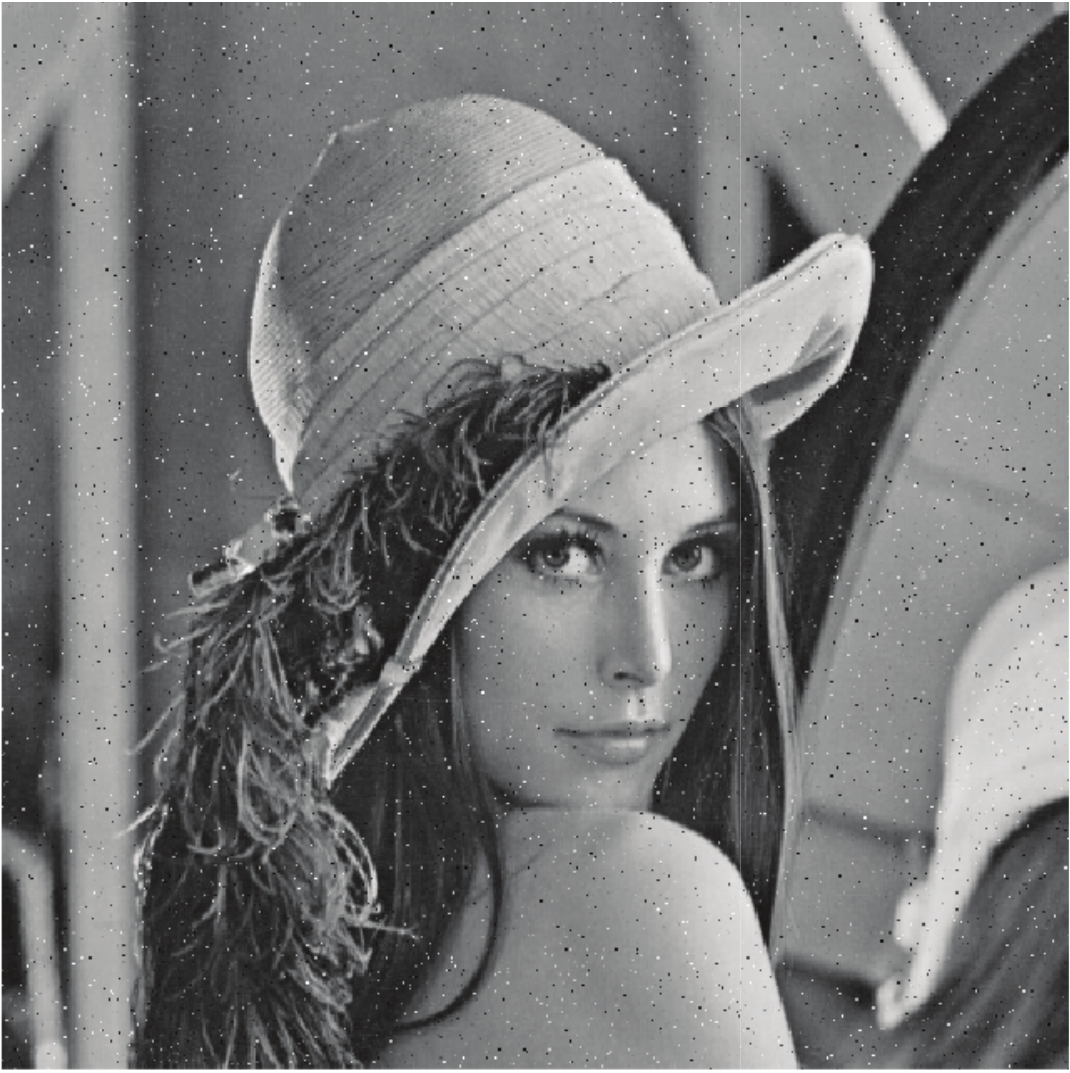}}
  \end{minipage}%
  \begin{minipage}[!t]{0.5\linewidth}
 \centering
  \subfigure[ ]{
    \label{Fig_exper_CDMA} %% label for second subfigure
    \includegraphics[scale=0.35,trim= 0 0 0 0]{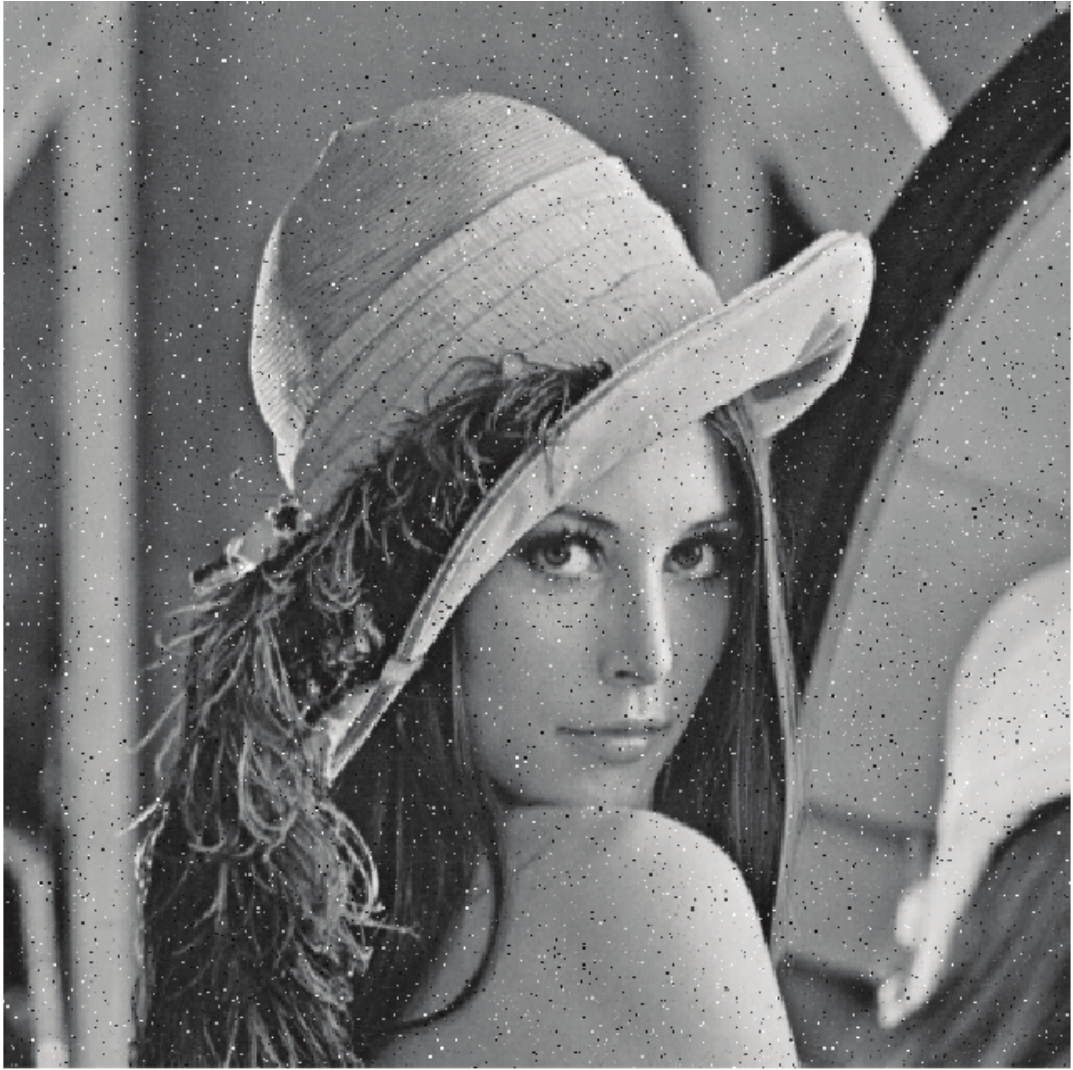}}
    \end{minipage}%
    \caption{The experiment results comparison for the CPOCMA and its competitors. (a) The origin image to be transmitted; (b) the received image decoded by CPOCMA; (c) the received image decoded by FDMA; (d) the received image decoded by CDMA.}
  \label{Fig_exper_result_photo} %% label for entire figure
\end{figure}
In order to verify the system performance of the CPOCMA in a practical environment, a communication system is established using two Wireless open-Access Research Platforms (WARPs) with Virtex-6 LX240T FPGA \cite{2020Wireless}. In each WARP, two MAX2820 RF chips can be used to build a dual-channel and dual-band transceiver in the 2.4GHz and 5GHz frequency bands with the maximum power gain 20 dBm. It can be used not only as the antenna at the transmitter end, but also for single-user/multiusers data reception at the receiver. The experiment system configuration is shown in Fig. \ref{Fig_setup_photo}, where the Laptop is used to generate the transmitted signal and decode the received signal. A data frame contains a Start of Frame Delimiter (SFD) and the information signal encoded using different comparison methods sequentially, as shown in Fig. \ref{Fig_frame_str}. SFD is a 128 training bits sequence generated by Golden code sequence. The information signals in each frame carry 512 bits for each individual method with sampling frequency 40MHz (totally 512*3 = 1536 bits). The frames are sent to the WARP at the transmitter through the Ethernet port. The test frames are radiated by an antenna and transmitted through the real radio channel. The single-user/multiusers (maximum two users for one WARP) at the receiver also use the Ethernet to send the received signal to the Laptop computers to demodulate the information. The SFD is used at the receiver to perform the frame synchronization and frequency offset compensation task.

Figure \ref{Fig_exper_result_photo} shows the image transmission results using the proposed CPOCMA, the CDMA and the FDMA in the same scenario using the same transmitter power 8dBm, where Fig. \ref{Fig_exper_orig} is the original image to be transmitted, Figs. \ref{Fig_exper_CPOCMA}, \ref{Fig_exper_FDMA} and \ref{Fig_exper_CDMA} are the received images using the CPOCMA, FDMA and CDMA with subcarriers $N$=2, respectively. It can be found that the received image in CPOCMA system shows higher quality with less distortion, less salt and pepper noise as compared to those using the FDMA and the CDMA. The Peak Signal to Noise Ratio (PSNR) is used as the criterion to evaluate images quality, where the corresponding PSNR of the four images is given in Table \ref{tab_PSNR}.  From Table \ref{tab_PSNR}, we see that the PSNR of CDMA shows the minimum value, which means that the greater distortion caused by the transmission bits error. It is consistent with the observation by naked eye and the simulation results.

\begin{table} \centering
\caption{\label{tab_PSNR}The PSNR of the origin image and the decoded images.}
%\begin{ruledtabular}
%\begin{spacing}{1.2}
\begin{tabular}{cccccc}
  \hline
   & Origin image & CPOCMA & FDMA & CDMA \\
  \hline
  PSNR & 32.2630 & 29.9408 & 27.7157 & 26.5614 \\
  \hline
\end{tabular}
%\end{spacing}
%\end{ruledtabular}
\end{table}

\begin{figure*}
\begin{minipage}[!t]{0.5\linewidth}
\centering
  \subfigure[ ]{
    \label{Fig_exp_BER} %% label for first subfigure
    \includegraphics[scale=0.55,trim= 0 0 0 0]{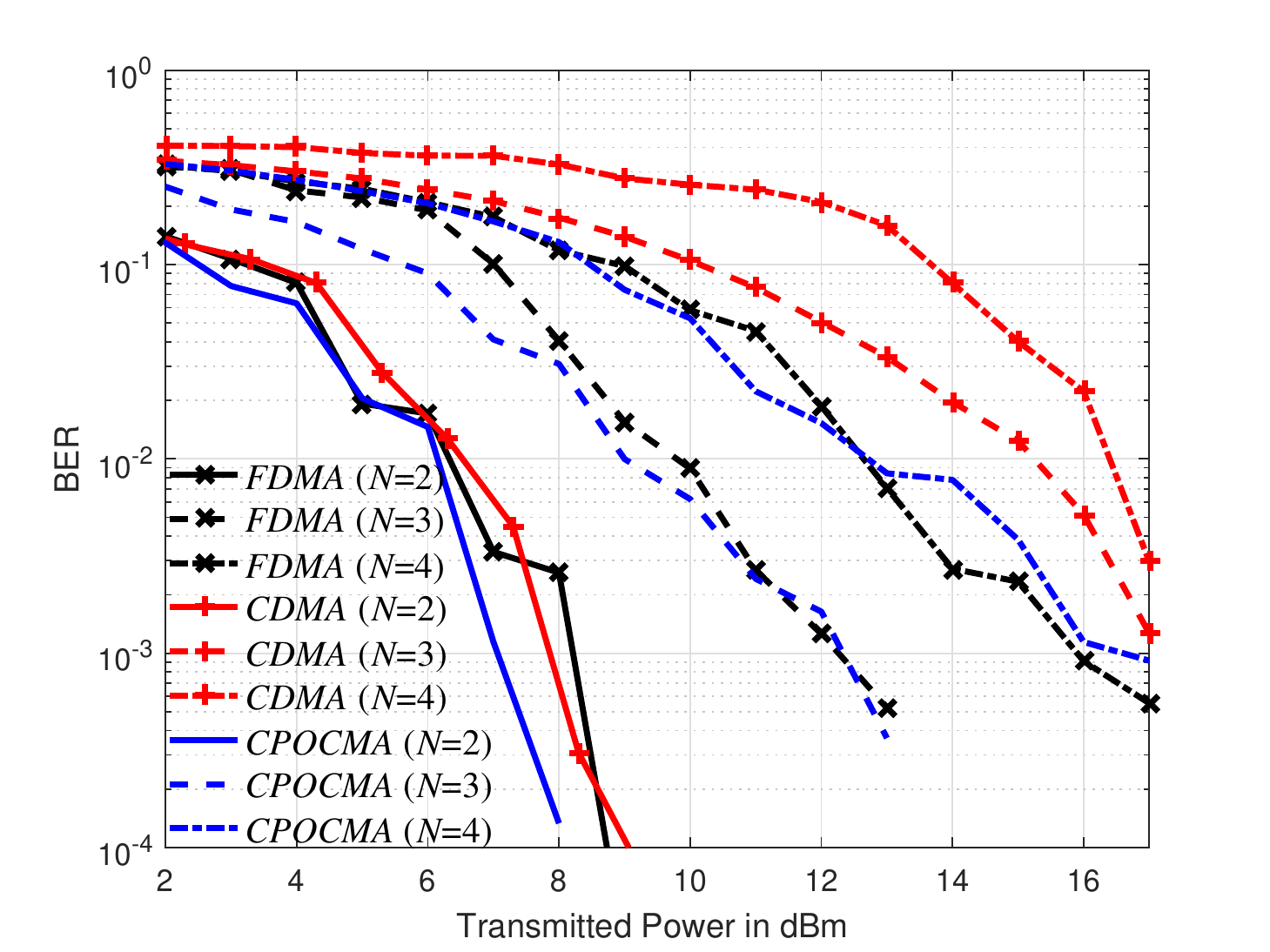}}
  \end{minipage}%
  \begin{minipage}[!t]{0.5\linewidth}
 \centering
  \subfigure[ ]{
    \label{Fig_exp_est} %% label for second subfigure
    \includegraphics[scale=0.55,trim= 0 0 0 0]{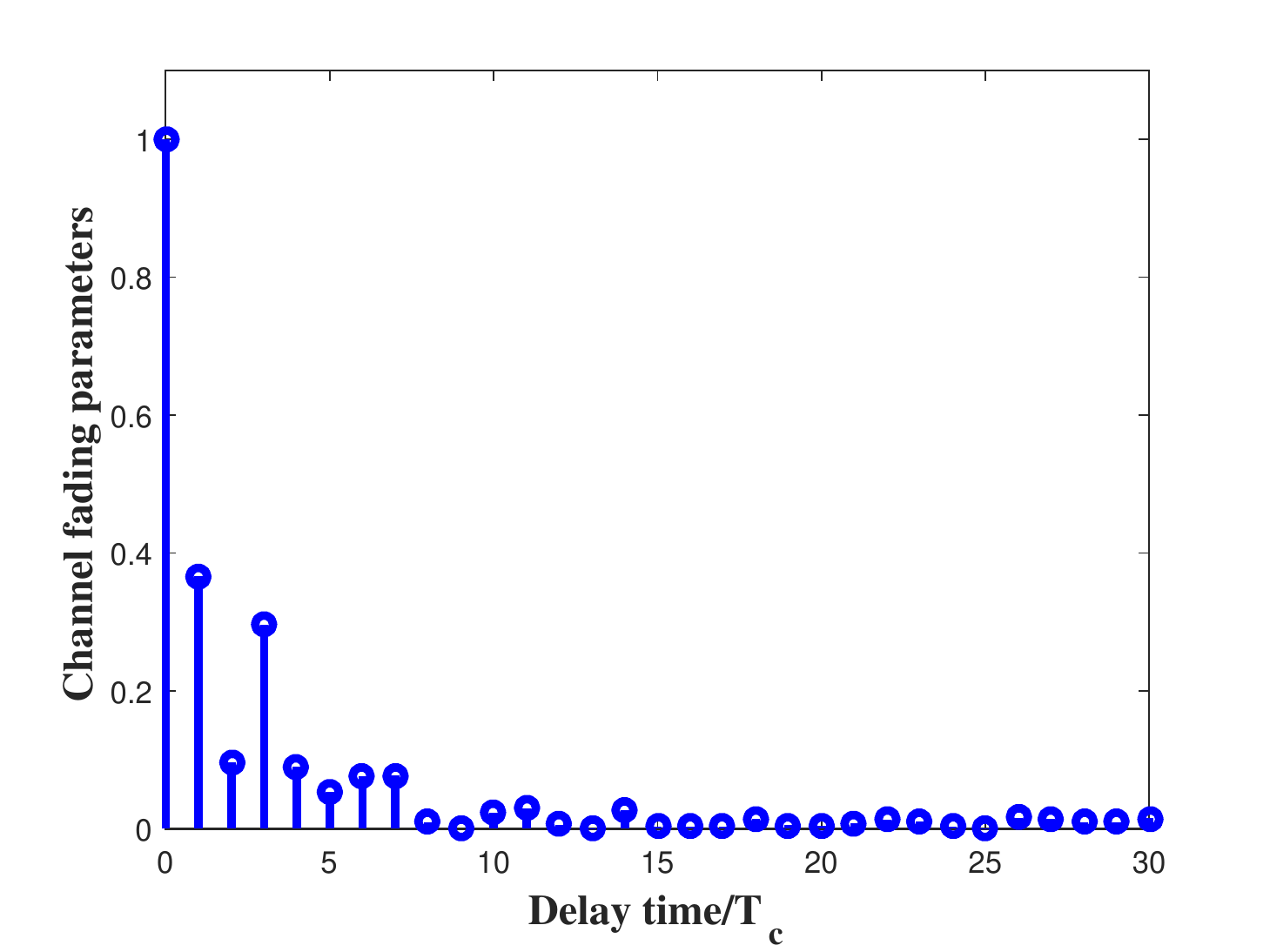}}
    \end{minipage}%
    \\
    \caption{The experimental BER for the single-user case with $N$ = 2, 3, 4. (a) The experiment BER result; (b) the estimated channel parameters.}
  \label{Fig_exp_single} %% label for entire figure
\end{figure*}

\begin{figure*}
\begin{minipage}[!t]{0.5\linewidth}
\centering
  \subfigure[ ]{
    \label{Fig_exp_BER_multi} %% label for first subfigure
    \includegraphics[scale=0.55,trim= 0 0 0 0]{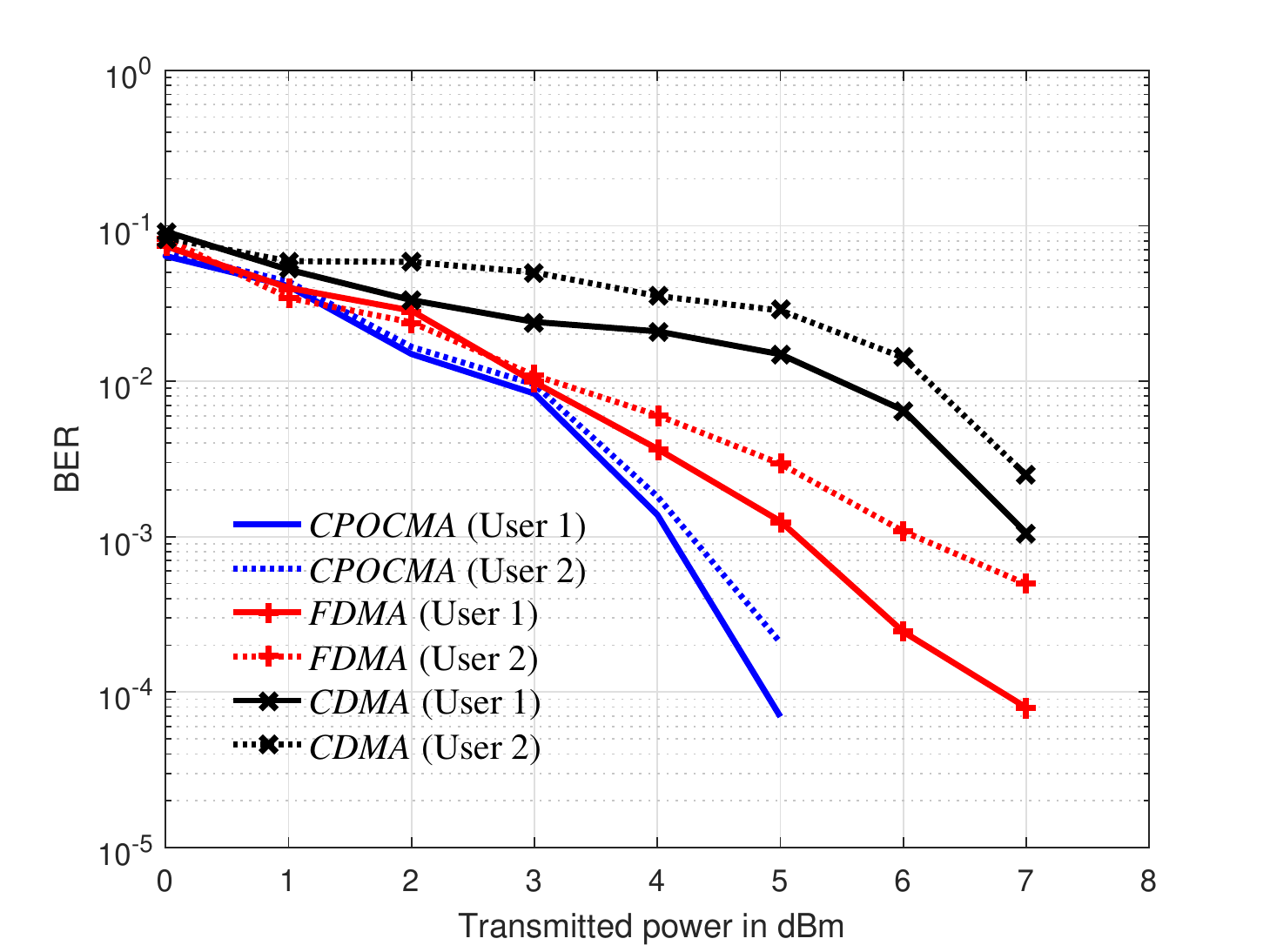}}
  \end{minipage}%
  \begin{minipage}[!t]{0.5\linewidth}
 \centering
  \subfigure[ ]{
    \label{Fig_exp_est_multi} %% label for second subfigure
    \includegraphics[scale=0.55,trim= 0 0 0 0]{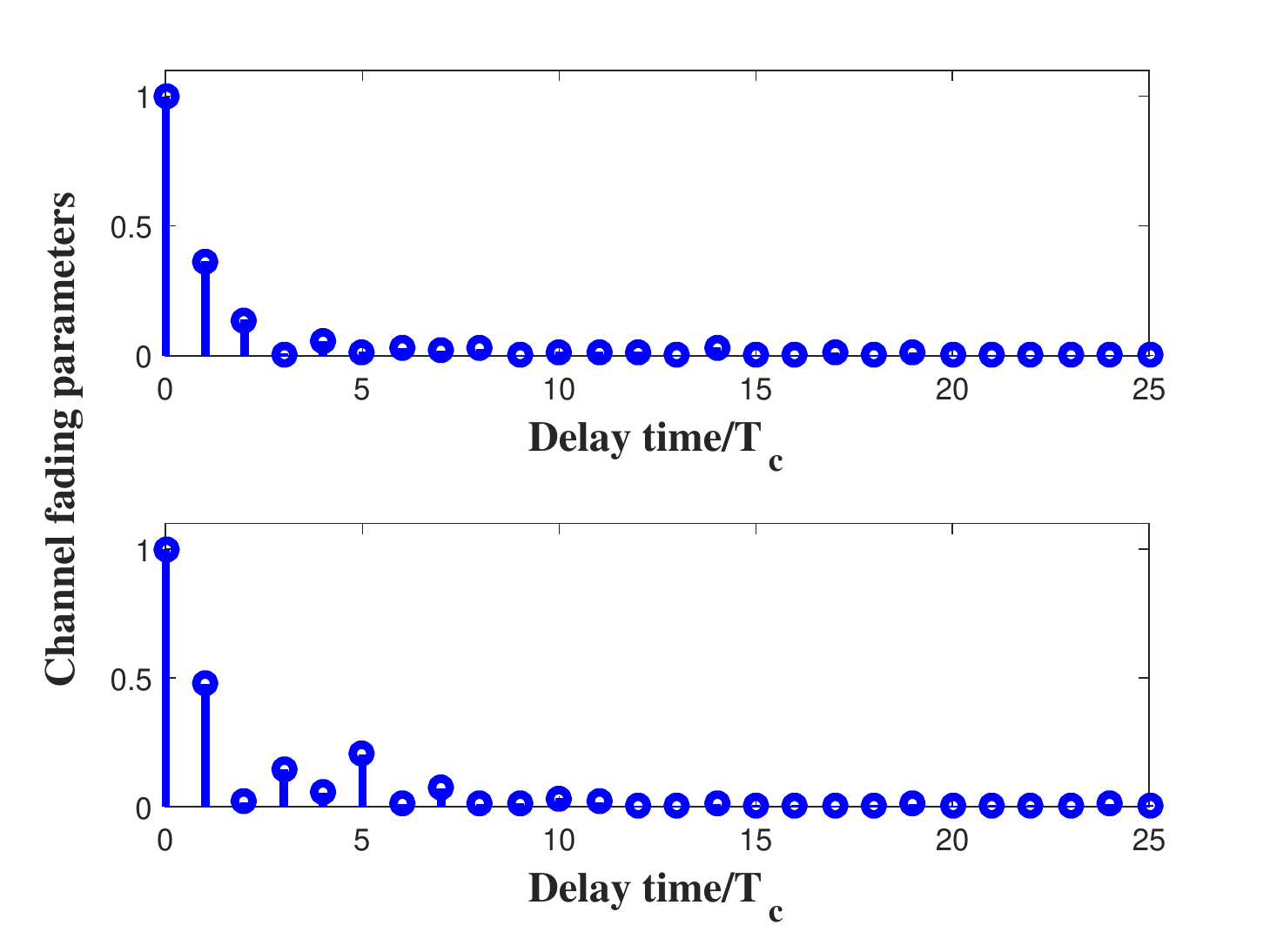}}
    \end{minipage}%
    \\
    \caption{The experimental BER for the multi-user case in practical radio channel. (a) The experiment BER comparison result; (b) the estimated channel parameters of User 1 and User 2 are given in the upper panel and the lower panel, respectively.}
  \label{Fig_exp_multi} %% label for entire figure
\end{figure*}

Figure \ref{Fig_exp_BER} shows the BER curves versus the transmission power of the proposed CPOCMA, FDMA and CDMA system over the three-paths channel in the out-door environment, where the transmission power is adjusted to simulate the SNR change in the practical experimental environment. The BER performances of the single-user with various subcarriers $N$ = 2, 3, 4, respectively, are shown in Fig. \ref{Fig_exp_BER}. It can be found that the performance of CPOCMA is slightly better than that of FDMA and CDMA, the trend of the experiment results is consistent with that of the simulations. The corresponding channel parameters estimation is performed using Least Squares algorithm, the result is given in Fig. \ref{Fig_exp_est}, where the delay unit ${T_c} = {\rm{ 3}}.{\rm{2}}\mu s$.

Since the receiver WARP hardware with two antennas supports maximum two users, the multi-user experiment is conducted. In this configuration, the transmitted signals are generated at the same transmitter, then received by the individual antenna of two users, respectively, as shown in Fig. \ref{Fig_setup_photo}. The experimental BER comparison for the same date transmission rate of the proposed CPOCMA, FDMA and CDMA are shown in Fig. \ref{Fig_exp_BER_multi}, and the corresponding channel estimation of two users are given in the upper panel and the lower panel of Fig. \ref{Fig_exp_est_multi}, respectively. User 1 and User 2 of CPOCMA show better BER than those of FDMA and CDMA, especially, showing a significant performance improvement under the high transmission power (approximately equivalent SNR). The results verify the feasibility and superiority of the proposed CPOCMA.

\section{Conclusion}
A chaotic pseudo-orthogonal carrier multi-access communication system is presented in this work. A CPOSF bank is proposed to generate multi-carrier signals encoding information bits. The transmitted signals share the same center frequency with limited bandwidth to improve the energy efficiency. At the receiver, the CPOCF bank and the CPOMF bank are used together to extract the subcarriers from the received signal and reduce the effect of noise from the corresponding subcarrier, respectively. It effectively reduces the MAI caused by the additional subcarriers, improves the communications system performance under the multiuser case. The performance of the proposed system is analyzed and BER expression for AWGN channel is derived. The simulation results show that the BER performance of the CPOCMA outperforms the CSF and CCBFM in the single-user case, and outperforms the FDMA and CDMA in multiuser case. Moreover, the throughput per frequency of CPOCMA is significantly higher than those of CDMA and FDMA, which is a most significant feature needed by IoT applications. An experimental communications system based on WARP is designed to test the communication performances of the CPOCMA, FDMA and CDMA schemes. The received image in CPOCMA shows higher quality with PSNR as compared to those obtained using FDMA and CDMA schemes under the same environment conditions and parameter configuration. Considering the demand of future IoT local area network wireless communications to multiuser at minimized bandwidth and energy costs, the proposed scheme shows the better potential in this scenario. The proposed method could be used together with the channel encoding methods like OFDM to further improve the data transmission rate in the future.

\section*{Acknowledgement(s)}

This research has been supported in part by China Postdoctoral Science Foundation Funded Project (2020M673349) and Open Research Fund from Shaanxi Key Laboratory of Complex System Control and Intelligent Information Processing (2020CP02).
%


\end{document}